\documentclass[useAMS,usenatbib, referee]{mn2e}
\usepackage{graphicx}
 \usepackage{times}
 \usepackage{rotating}
\usepackage{txfonts}

\def\vaa{\hbox{OGLE004336.91-732637.7}}
\def\vbb{\hbox{OGLE004633.76-731204.3}}
\def\vcc{\hbox{OGLE005100.18-725303.0}}
\def\va{\hbox{SMC-SC3}}
\def\vb{\hbox{SMC-SC4}}
 \def\gtrsim{\mathrel{\hbox{\rlap{\hbox{\lower4pt\hbox{$\sim$}}}\hbox{$>$}}}}
 \def\ltsim{\mathrel{\hbox{\rlap{\hbox{\lower4pt\hbox{$\sim$}}}\hbox{$<$}}}}

\title[High dispersion spectroscopy of two A supergiant systems in the SMC with novel properties ]{High dispersion spectroscopy of two A supergiant systems in the Small 
Magellanic Cloud with novel properties }

\author[Mennickent \& Smith]
  {R.E. Mennickent,$^1$\thanks{E-mail: rmennick@astro-udec.cl.
 Based on observations carried out at ESO telescopes: ESO proposal 69.D-0391(A), backup targets.
 }
  M. A. Smith$^{2}$\\
  $^1$Universidad de Concepci\'on, Departamento de Astronom\'{\i}a,
      Casilla 160-C, Concepci\'on, Chile\\
  $^{2}$  Department of Physics, Catholic University of America, Washington, DC 20064, USA; Present
address: Space Telescope Science Institute, 3700\\ San Martin Dr., Baltimore, MD 21218, USA }  
\date{}

\pagerange{\pageref{firstpage}--\pageref{lastpage}} \pubyear{2007}

\def\LaTeX{L\kern-.36em\raise.3ex\hbox{a}\kern-.15em
    T\kern-.1667em\lower.7ex\hbox{E}\kern-.125emX}

\begin{document}

\label{firstpage}

\maketitle

\begin{abstract}

 We present the results of a spectroscopic investigation of two novel 
variable bright blue stars in the SMC, OGLE004336.91-732637.7 (\va) and the 
periodically occulted star \vbb~ (\vb), whose photometric properties were
reported by Mennickent et al. (2010). 
High-resolution spectra in the optical and far-UV show that 
both objects are actually A + B type binaries.
Three spectra of \vb~ show radial velocity variations, consistent
with the photometric period of 184.26 days found in Mennickent et al. 
2010.  The optical 
spectra of the metallic lines in both systems show combined absorption and 
emission components that imply that they are formed in a flattened
envelope.  A comparison of the radial
velocity variations in \vb~ and the separation of the $V$ and $R$ emission 
components in the H$_{\alpha}$ emission profile indicate that this 
envelope, and probably also the envelope around \va, is a {\it circumbinary} 
disk with a characteristic orbital
radius some three times the radius of the binary system.  
The optical spectra of \va~ and \vb~ show, respectively, He\,I emission lines
and discrete Blue Absorption Components (``BACs") in metallic lines.
The high excitations of the He\,I lines in the \va~ spectrum and the 
complicated variations of Fe\,II emission and absorption components with
orbital phase in the spectrum of \vb~ suggests that shocks occur between 
the winds and various static regions of the stars' co-rotating binary-disk 
complexes.  We suggest that BACs
arise from wind shocks from the A star impacting the circumbinary disk
and a stream of former wind-efflux from the B star accreting onto the A 
star. The latter picture is broadly similar to mass transfer occurring in
the more evolved (but less massive)  Algol (B/A + K) systems, except that 
we envision transfer occurring in the other direction and not through the inner
Lagrangian point.   Accordingly, we dub these
objects prototype of a small group of Magellanic Cloud wind-interacting
A + B binaries.
\end{abstract}

\begin{keywords}
stars: early-type, stars: evolution, stars: mass-loss, stars: Ae
stars: variables-others
\end{keywords}

\section{Introduction}

   Mennickent et al. (2002, hereafter M02) have reported the existence of
a number of bright  blue stars in the Small Magellanic Cloud the light
curves of which exhibit periodic or quasi-periodic variability in their 
OGLE (Udalski et al. 1997) $I$-band light curves.  In an initial follow up
of the investigation, Mennickent et al. (2006; ``M06") noted that some stars 
exhibit peculiar spectroscopic properties and multiple periods.  They 
gave initial estimates of spectral classifications based on low dispersion 
spectra, but they noted conflicting properties.
In a companion study to this one, Mennickent et al. (2010; hereafter ``M10") 
have selected two novel Type\,3 variables\footnote{ Mennickent Type-3 
variables are SMC Be star candidates with $I$-band light curves varying 
periodically or quasi-periodically.}
with novel properties named
\vaa~  ($\equiv$ SMC-SC3-63371, MACHO ID 213.15560; hereafter \va) and 
\vbb~ ($\equiv$ SMC-SC4-67145, MACHO ID 212.15735.6; hereafter \vb).  
These Type\,3 stars were chosen according to the sole additional arbitrary 
criterion that they are brighter than $m_v$ = 14.2. 
The photometric analysis in M10
was based on new OGLE III  light curves and found that \va~ 
exhibits a double-minimum period with a period of 238.1 days and a
second quasi-period that fluctuates between two values each
near 15.35 days. The appearance of the longer
period, which exhibits minima at unequal spacings and unequal depths
suggests the appearance of an ellipsoidal variable. The high orbital 
eccentricity implied by these properties is confirmed by the
inference from the sharp and unequal light maxima that reflection
effects from a secondary occur during brief periastron passages.
The light curve of \vb~ reveals surprising periodic deep 
eclipses every 184.26 days, modulated over a superperiod of 
several cycles, and chaotic variability indicative of multiple 
shorter periods.

M06 and M10 have open a variety of questions  that can be pursued with
high resolution spectra, which we undertake to pursue now. We have already
remarked as one example that  the spectra of \va~ and \vb~ spectra are
unusual in that they show properties of rather conflicting spectral types
(strong Na\,D  and strong H$_{\alpha}$ and H$_{\beta}$ emission components).
Therefore, we obtained three
high-dispersion optical spectra of \va~ and \vb~ during the period 2002-9.
To obtain a better understanding  of how the spectra appear at
short wavelengths we obtained a {\it Far Ultraviolet
Spectroscopic Explorer} ({\it FUSE}) spectrum of both stars.
These spectra and the {\it FUSE} data form the basis of this investigation. 
The goal of our
study is to elucidate the properties of the circumstellar environment of 
these objects and to pave the way to understanding the evolutionary state 
of this potential new subclass of Ae stars.  




\begin{table*}
\centering
 \caption{Summary of observations. The MJD numbers at mid exposure are given.
The periods detected in photometric data are also given. Phases refer to the 
ephemeris given in Mennickent et al. (2010).  CD refers to cross-disperser.}
 \begin{tabular}{@{}cccccccccccc@{}}
object &periods (d) & instrument &UT-date   &    airmass &$\Delta\lambda$ (\AA)&   grating & exptime (s) &mjd-obs &phase &$S/N$\\
\hline
\va~ & 238, 15 &UVES &2002/05/17&  1.96&3100--8500 &CD\#1,3&  600 &52411.39633  &0.28&25\\
\va~ & &UVES &2002/05/17&  1.91&3100--10400  &CD\#1,4&  600 & 52411.40536 &0.28&20\\
\vb~ & 184 &UVES &2002/05/17&  1.86&3100--8500  &CD\#1,3&  600 &52411.41614 &0.23&20\\
\vb~ & &UVES &2002/05/17&  1.81&3100-10400  &CD\#1,4&  600 & 52411.42782 &0.23&15\\
\va~ & &MIKE  &2007/11/09&  1.43 &3390--9410 &echelle&  500 &54413.03654  &0.69   & 25        \\
\vb~ & &MIKE  &2007/11/09&  1.42&3390--9410 &echelle&  500 &54413.04421   &0.10    &  40      \\
\va~ & &Echelle  &2009/08/25  &  1.44 &3940--7490 &echelle&  2000 &55068.24680 &0.44   &      17 \\
\vb~ & &Echelle &2009/08/25&  1.40&3940--7490 &echelle&2000   &55068.27751    &0.66    &17       \\
 \hline
\end{tabular}
\end{table*}

\begin{table*}
\centering
 \caption{Summary of {\it FUSE} Observations.  Phases refer to the ephemeris given in Mennickent et al. (2010).}
 \begin{tabular}{@{}cccccc@{}}  
object & UT-date  &UT-start & exptime (s) &mjd-obs &phase\\
\hline
\va~   &  2006/10/05&04:05:06& 2430  &54013.18427083 &0.01\\
\vb~   &  2006/11/30&11:36:26& 2736  &54069.49946759 &0.24\\
 \hline
\end{tabular}
\end{table*}

\section{Observations and data reduction}

High resolution spectra  were obtained on 2002 May 17 with the ESO
Ultraviolet-Visual Echelle Spectrograph in dichroic modes at the UT\,2 
telescope in the ESO Paranal Observatory, Chile. The three CCD chips 
on this instrument allowed sampling the spectral range of 3100--10400 \AA.  
A slit width of 1" allowed us to obtain spectra at resolving power 
$\sim$ 40\,000.  
The spectra were normalized to the continuum, and no flux calibration 
or telluric correction was needed. 
Additional spectra were obtained with the {\it Magellan 
Inamori Kyocera Echelle} (MIKE) spectrograph at the Clay Telescope in 
Las Campanas Observatory, Chile on 2007 November 9. 
This double echelle spectrograph provided wavelength coverage of
3390$-$4965 \AA~(blue camera) and 4974$-$9407 \AA~ (red camera). 
With a slit width of 0.7" the resolving power was again 40\,000.
The spectra were reduced and calibrated with IRAF. 
We obtained a third spectrum for each star on 2009 August 25 with the 
echelle spectrograph mounted in the 2.5m DuPont telescope of Las Campanas 
Observatory, Chile; this is referred to as the ``LCO" spectrum in our
figure captions. The wavelength range in this case was 3940$-$7490 \AA~
and the resolving power again of 40\,000. These spectra were also reduced and 
calibrated with standard IRAF routines. 

A summary of the optical spectroscopic observations is given in Table\,1. 
The spectra of \vb\ were obtained at phases 0.23, 0.10, and 0.66, 
respectively, in the 184.26 day period identified in M10. 
The zeropoint is taken as the mid-eclipse in the light curves, i.e., at the
light minima. For \va\ the spectra were taken at phases 0.28, 0.69, and 0.44, 
according to the 238.1 day period identified in M10. 

In addition,  far-UV spectra of \va~ and \vb~ were obtained through 
the large science aperture of the {\it Far Ultraviolet Spectroscopic Explorer} 
({\it FUSE})  in 2006 October-November
under GO Cycle 6 Program F907, as detailed in Table\,2. 
A spectrum of \vcc\ was obtained on 2006 October 6, as a quasi-standard.  
The {\it FUSE} spectra 
cover the continuous wavelength range 929-1188 \AA\ 
over eight MAMA detector segments.  The {\it FUSE} spectrum is recorded 
on two of these segments, thereby insuring the
acquisition of two independent and simultaneous spectra. The spectral 
resolving power among these varies with wavelength but is typically in
the range 15\,000-20\,000.  These spectra were reduced
with the CalFUSE version 3.1.8 pipeline system, which was similar to 
version, v3.2, which was used for the final reprocessing of the {\it FUSE} 
archive.  Examination of auxiliary files that display the positions of
photon events on the detector in the spatial direction confirm that
they originate from an effective point source. This fact effectively
rules out that the ultraviolet fluxes are contaminated by a nearby
comparably bright source in the science aperture.

\section{Results}

The optical spectra of the two program stars are complicated, and it is 
therefore helpful to describe their general properties before carrying
out a quantitative analysis of their spectral features.
These spectra are redshifted by amounts of roughly 2 \AA, which is typical 
for optical spectra of members of the SMC.

\subsection{Reconnaissance of the optical spectra}

  In many respects the optical spectra are similar to the
spectrum of a typical A supergiant.   This statement is demonstrated in 
Figure\,1, which shows the high-level Balmer lines and the blue-green UVES 
spectra of \va~ and \vb~ and of the Galactic A5\,II star HD\,74252 from the 
UVES atlas (Bagnulo et al. 2003). The cores of the hydrogen and metallic 
lines in the spectra of the program stars are influenced 
by two competing effects.  On one hand, the strengths of the
metallic lines are weakened in our objects due to their low metallicities.
This effect is largely compensated by the presence of a substantial disk 
contribution, the details of which we present below.  
The cores of both hydrogen and metallic lines are stronger in \va~ than in 
\vb.~ Detailed inspection shows that most of these lines arise from once-ionized
Fe-like ions and of mixed stages of light metals (Mg\,I, Si\,II, Ti\,II, 
Ca\,I), and these will be discussed in $\S$3.5-3.7. Thus, the correspondence
is good between the optical spectra of our stars and a middle-A, class I-II 
supergiant. As noted in M10, an A supergiant spectral type is also 
inferred from the colors and magnitudes of these objects. 
The spectral types will be refined in our discussion of the Ca\,II line 
profile ($\S$3.2).

\begin{figure}
\centering
 \scalebox{1}[1]{\includegraphics[angle=90,width=8cm]{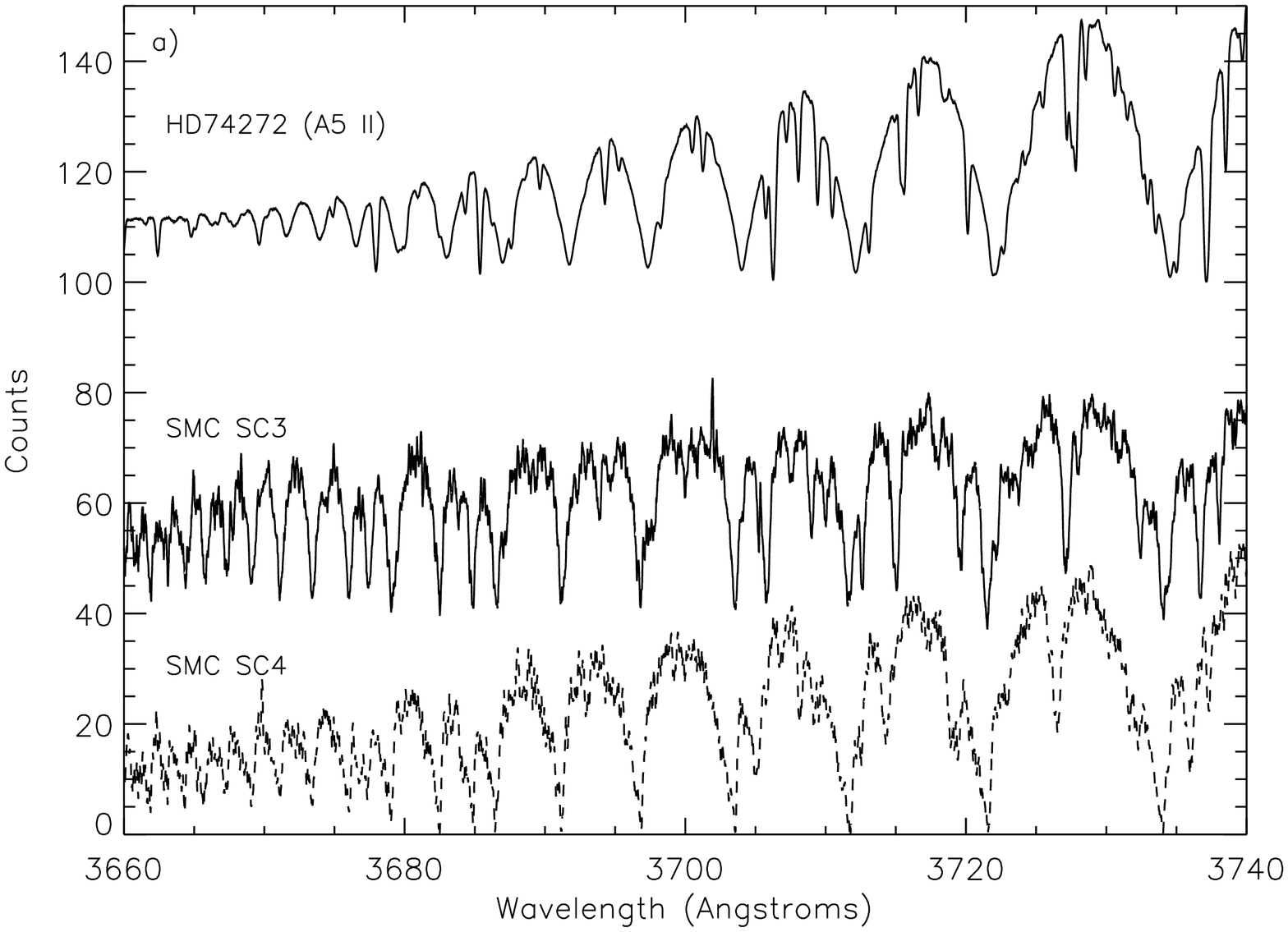}}
 \scalebox{1}[1]{\includegraphics[angle=90,width=8cm]{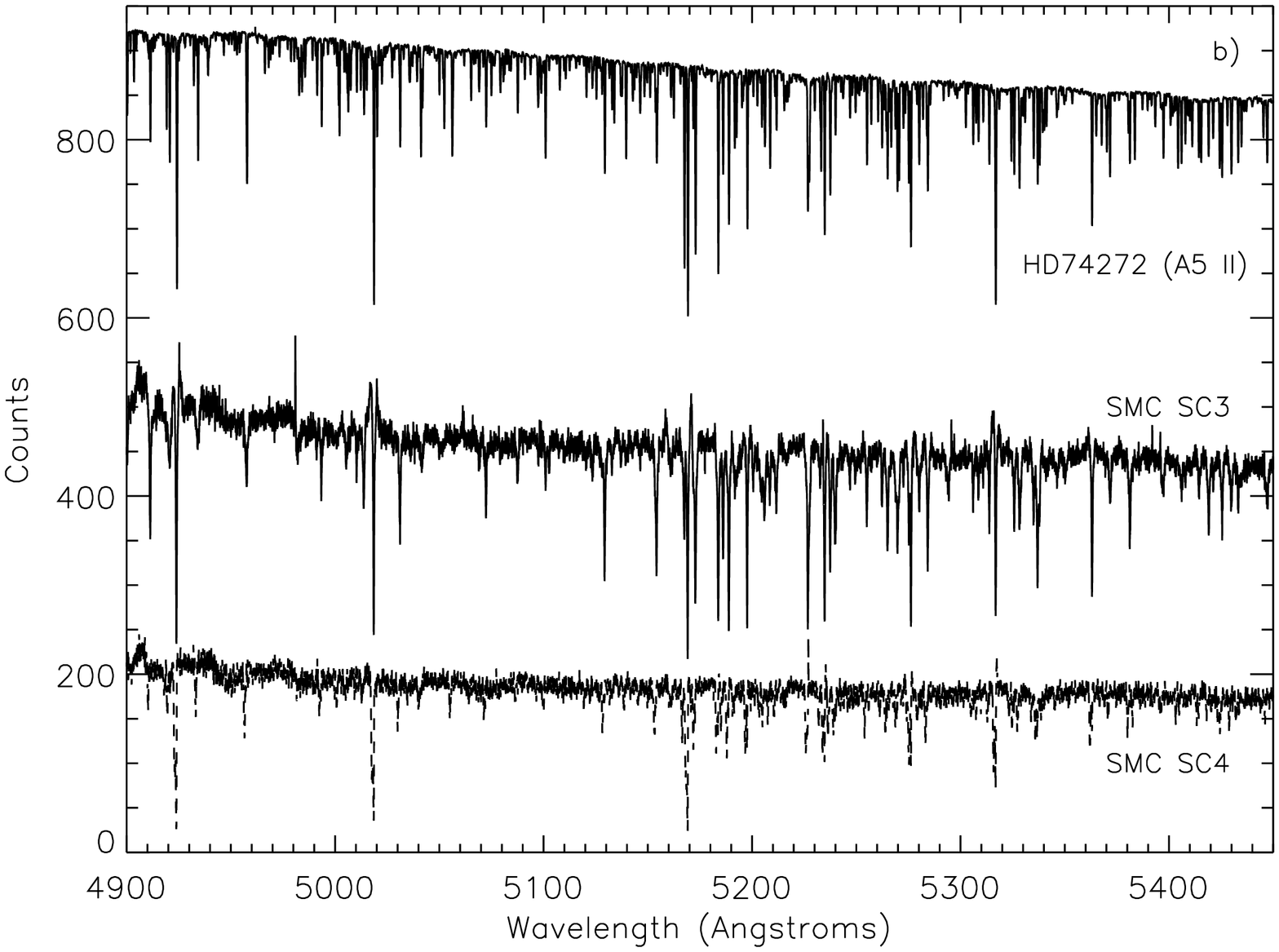}}
\caption {The optical UVES spectrum of the target stars and the Galactic 
A5II standard HD\,74252 taken from the UVES atlas covering (left panel) the 
high level Balmer
lines and (right panel) the metallic lines in the region 4900-5450\,\AA.~ 
The standard and \va~ have been offset in flux for convenience, and \va~ and
\vb~ have been blueshifted to the rest frame. The general appearance of the
metallic lines shown for the two SC targets is similar to that of the standard's
lines. The Fe\,II lines 4923\,\AA\, 5018\,\AA\, 5176\,\AA\, and 5316\,\AA, are
discussed in the text. Careful inspection shows that the wings of these 
particular lines show emission in \va.}
\label{1}
\end{figure}

 However, in other respects there are stark contrasts between the spectra 
of our program stars and A supergiants. We have already pointed out  
some of the strongest lines in our optical spectra, such as 
the Na\,I doublet, (Figure\,2) and the 
strong H$_{\alpha}$ emission.
In contrast to the merely faint Beals Type\,I P\,Cygni emission, 
e.g., in the Bagnulo et al. (2003) spectrum of the A6\,Ia standard
HD\,97534 the  H$_{\alpha}$ profiles are not only strong but double peaked
with nearly equal ``V" and ``R" emission components - see Figure 3. 
The strengths of these components are shown in Table 3.
These H$_{\alpha}$ profiles also resemble those of Be stars at the high
end of their distribution of H$_{\alpha}$ strengths,  although they
are not quite extraordinary among Be stars. 
These attributes indicate the presence of a flattened, 
Keplerian disk.  In contrast to the \vb~ profiles, the H$_{\alpha}$
emission of \va~ decreased over time, suggesting that these changes
were monotonic over time and not due to stellar or binary activity. 

\begin{figure}
\centering
 \scalebox{1}[1]{\includegraphics[angle=90,width=8cm]{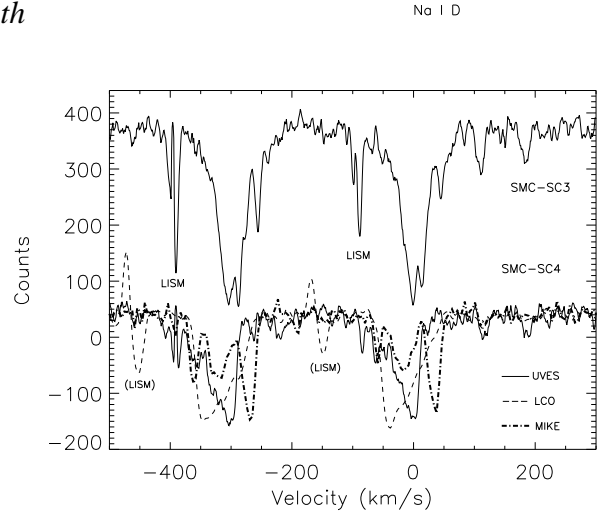}}
\caption{A comparison of the Na\,I D doublet 
lines in our UVES spectra, smoothed over 4 points.
These systems are placed together to show the relatively symmetric
profiles for \va~ and the great variations for the profiles of \vb.~
In the latter star the blue/red wings appear depressed in the 
 2002 and 2009 spectra, respectively, but the MIKE 2007 spectra shows 
that these shadings are caused by blendings of BAC subcomponents of 
widely varying relative strengths. 
The zeropoints of the velocity system are referred to the rest frames 
of the D2 (5896\,\AA), and in so doing the various radial velocities of
the primary star have been removed. The counts for the SMC-SC4 spectra
 have been displaced downward by 150 units for convenience. 
The emission features in the LCO spectra of \vb~ are telluric D line emission.  
}
\label{2}
\end{figure}

    The optical spectrum of SMC-SC3 is likewise well populated with sharp 
lines of light and Fe-group elements.  The Fe\,II lines, those arising from 
levels at 2.9\,eV are particularly prominent. All of these are characteristic 
of the spectrum of a Be star disk seen at high inclination.
In a few cases discussed below weak symmetric emissions
are superposed on some of the strongest Fe\,II lines in the optical spectrum.
However, unlike the spectra 
of B[e] stars, lines arising from metastable levels are not present, 
nor are [Fe] emission lines.  Since  these are a defining traits of 
B[e] stars,  this star cannot be a B[e] (or ``A[e]") star.

\begin{figure}
\centering
\scalebox{1}[1]{\includegraphics[angle=-90,width=8.5cm]{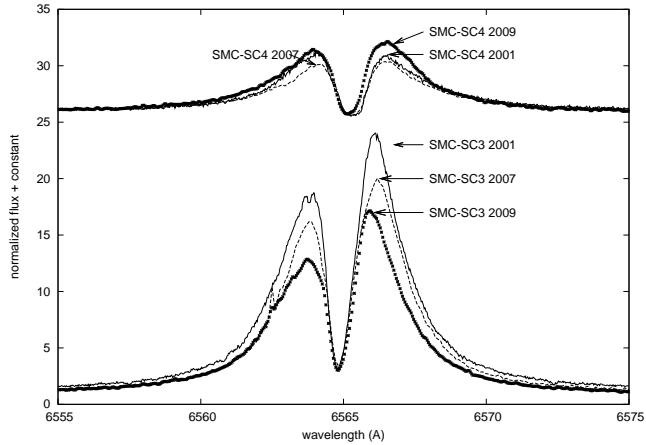}}   
\caption{The UVES (2002), MIKE (2007) and LCO (2009) spectra depicting the 
H$_{\alpha}$ profile for both the program stars. Fluxes are normalized to 
a unit continuum level (with an offset of -25 for \vb), 
and heliocentric corrections have been applied. 
}
\label{3}
\end{figure}

\begin{table}
\centering
 \caption{Equivalent widths ($EW$), maximum intensity relative to the continuum ($I/I_{c}$) and
 full width at half maximum ($FWHM$) for the H$_{\alpha}$ emission line in 2002, 2007, and 2009.}
 \begin{tabular}{@{}cccc@{}}
  \hline
Object & -$EW$ (\AA) & $I/I_{c}$ & $FWHM$  (\AA)   \\
\hline
\va~  & 112-104-72 &24.1-22.7-17.1  &4.39-4.46-4.41  \\
\vb~  & 28-26-33   &6.3-5.7-7.1    &5.16-5.25-5.02  \\
\hline
\end{tabular}
\end{table}

In general the fitting of disk components to the metallic lines can be
performed to compute rough column densities by assuming a 
local disk temperature.
The spectrum of SMC-SC4 differs from the case of SMC-SC3, and for that 
matter from the spectrum of nearly all other known A or B stars, by 
the additional presence of a system of blueshifted components for
most metallic lines in the optical spectrum. 
These discrete Blueshifted Absorption Components, which we will refer 
to as ``BACs," are displaced by about -50 km\,s$^{-1}$ from what we will 
call the main (or red) sharp component, the velocities of which 
adhere closely to those of the hydrogen line cores in the spectrum. Among the
metallic lines the BACs are generally of comparable strength and often
stronger than the main components. They tend to be barely visible or 
absent in weak resonance lines, e.g., in the region 3820-3850\,\AA.~ 
In the strong Fe\,II lines and the Na\,I D doublet a secondary BAC is also
present at approximately -100 km\,s$^{-1}$ relative to the main component.
Section 3.7.1 will be devoted to a discussion of these novel features.

\subsection{Spectral Types from the Ca\,II H/K features }

  Despite the comingling of photospheric and disk features in many lines
of the optical
spectrum, a spectral type for our stars can be determined
from the Ca\,H or K lines (with correction for metallicity)
and the wings of the hydrogen lines. The high level Balmer
and Paschen lines are consistent with a middle A-type spectral type.
However, these features can vary significantly among spectra of A-type 
standards separated in type or luminosity class. As already mentioned,
the emission in the lower members also interferes with a comparison to
profiles from spectral standards or synthesis models. Consequently, we found
the Ca\,II K line to be a more reliable indicator of the positions of the
stars in the HR Diagram.  The wings of this line are sensitive to
electron pressure and thus luminosity class. The hydrogen lines and Fe-line
metal lines already guide the spectral classification to the middle A range.

  Figure\,4 exhibits matches we have found for the K lines of SMC-SC3 and 
SMC-SC4 with the spectral standards HD\,97534 (A6\,Ia) and HD\,74272 (A5\,II)
from the UVES atlas, again from the UVES atlas. We have likewise
fit the spectra with {\sc SYNSPEC} models for T$_{\rm eff}$ = 8500\,K
and log\,g = 3.  The explicit assumption we made in our fittings
was that there is no measurable contribution at this wavelength from
another continuum source such as a secondary star.  In general, the fits 
are very good, except that the models predict a narrower core than is
observed in SMC-SC3 or even the core of the Ia standard (Fig\,4a). 
In addition, the blue core of both the SMC-04 and class two standard
shows an apparent ``emission" feature" that the model atmospheres 
(with no chromosphere) do not produce (Fig.\,4b). Otherwise, we
estimate the precision to be ${\pm 1}$ spectral subtype,
${\pm 500 K}$  in T$_{\rm eff}$  and ${\pm 0.5}$ dex in log\,g. 
In view of this self-consistency of our fittings with the position of the
stars in the HR Diagram (M10), we judge the principal features of the 
optical spectrum including the Ca\,K  and hydrogen lines to be due to middle
A supergiants, specifically types near A6\,I and A5\,II, respectively.
In addition  we judge that we should just be able to discern a 6\% (3
magnitudes) contamination to the K line from flux of a contributing secondary. 


\begin{figure}
\centering
\scalebox{1}[1]{\includegraphics[angle=90,width=8cm]{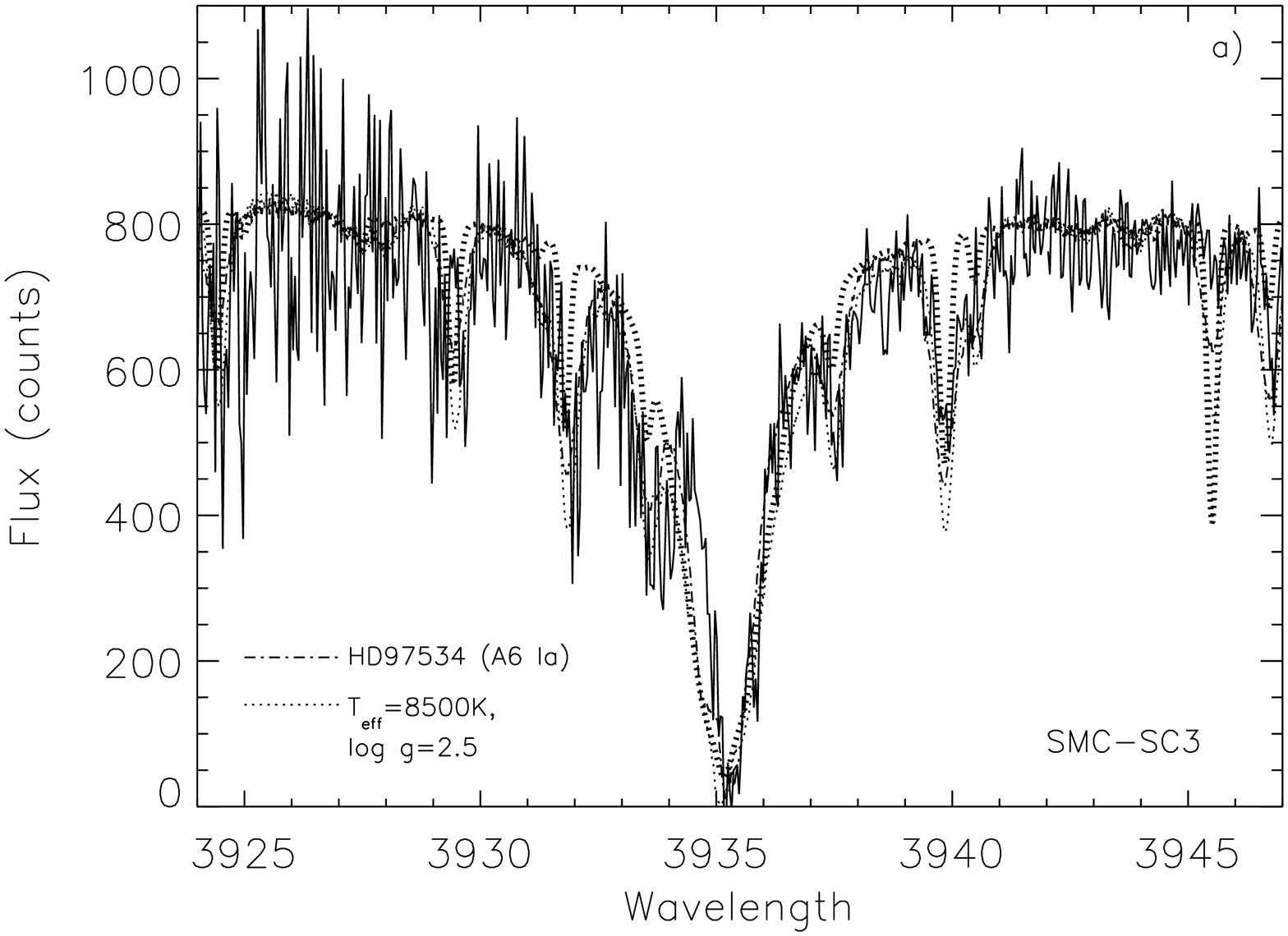}}
\scalebox{1}[1]{\includegraphics[angle=90,width=8cm]{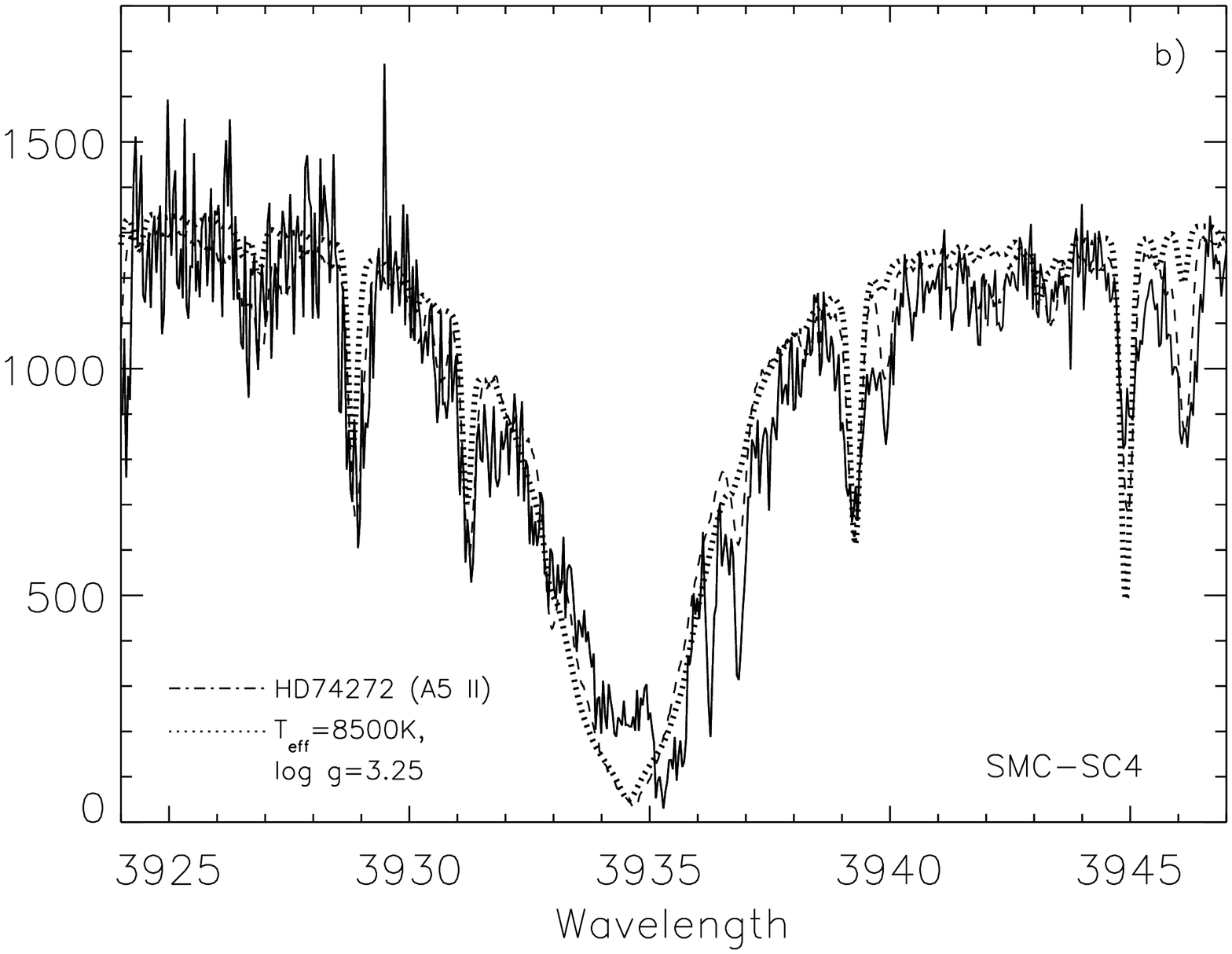}}
\caption{
The spectral region surrounding the Ca\,II K line in our two
program stars from our MIKE (2007) spectra. Spectra of the
comparison  stars HD\,97534 and HD\,74272 are represented by dashed
lines. Fits from {\sc SYNSPEC} models using Kurucz atmospheres
are given by the dotted lines.
}
\label{4}
\end{figure}

\begin{figure}
\centering
\scalebox{1}[1]{\includegraphics[angle=90,width=8cm]{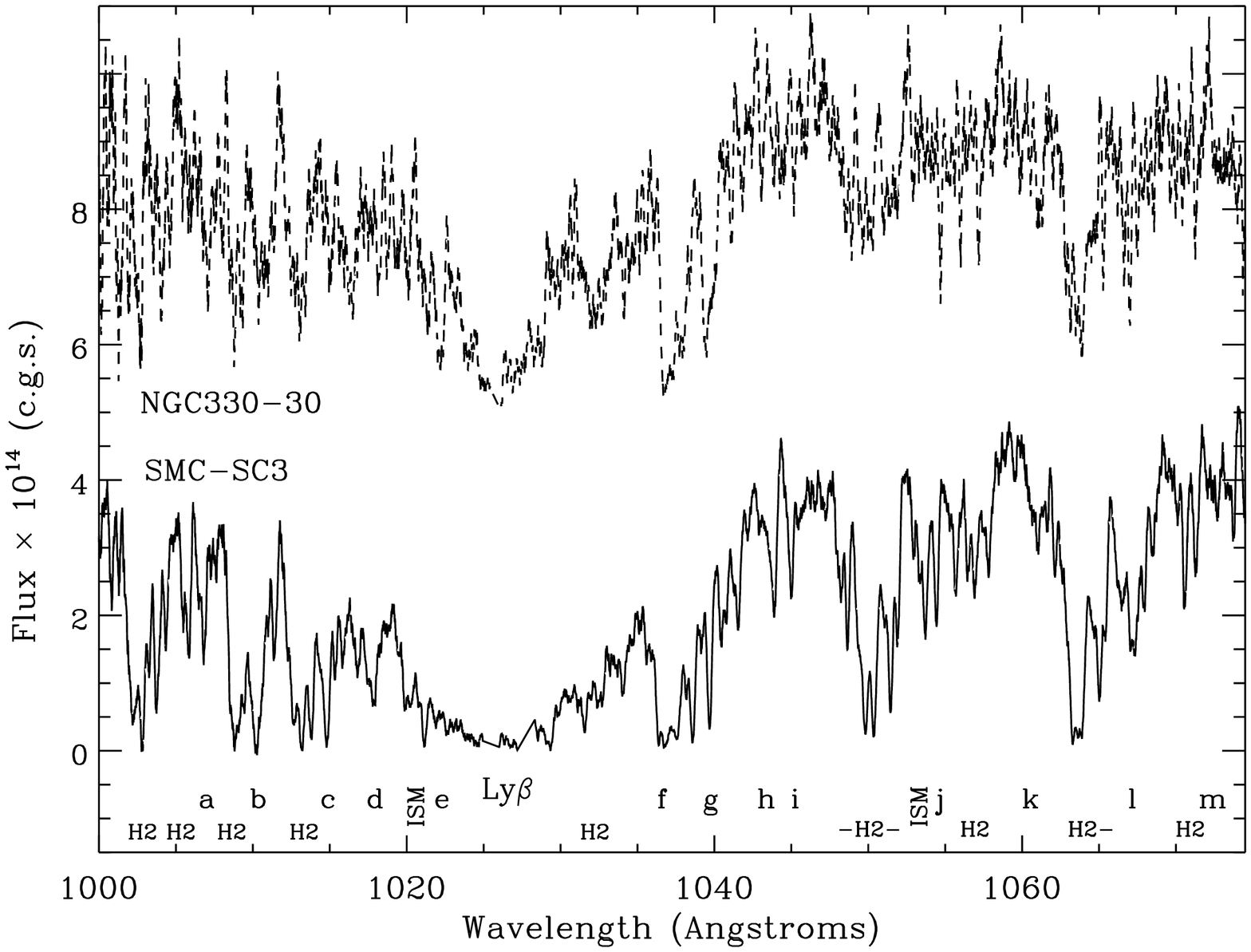}}
\scalebox{1}[1]{\includegraphics[angle=90,width=8cm]{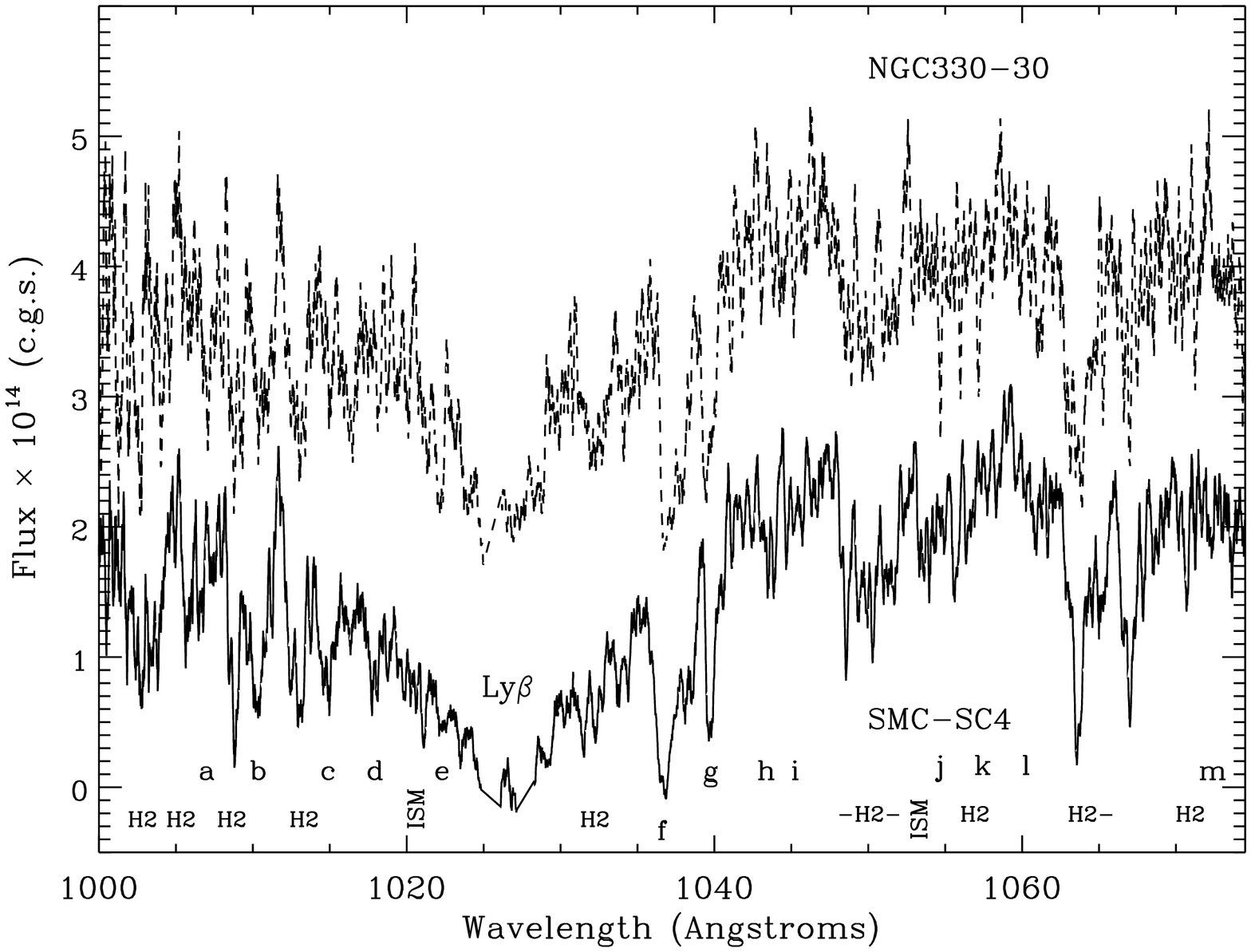}}
\caption{ The far UV spectrum covering 1000-1070\,\AA~ for the B3\,II 
quasi-standard of NGC\,330-30  and one of our program stars, 
smoothed over 8 points.~ The spectrum 
of the standard is offset vertically for clarity.  Several molecular H$_2$ 
and atomic interstellar (``ISM") lines are noted. We identify photospheric 
lines by coded letters. These codes have the following meaning:~
a) Fe\,III-Cr\,III 1007\AA,~ b) C\,II 1010.4\,\AA,~ c) S\,III 1015.5\,\AA,~
d) Fe\,III 1018.4\,\AA,~ e) Fe\,III 1021.7\,\AA,~ f) C\,II 1036-1037\,\AA,~
g) Fe\,III 1039.9\,\AA,~ h) Cr\,III 1042.9\,\AA,~ i) Fe\,III 1045.2\,\AA,~
j) Mn\,III 1055.5\,\AA,~ k) Fe\,III 1058.8\,\AA,~ k) Fe\,III 1060.7\,\AA,~ 
l) Si\,IV 1066\,6\,\AA,~ and m) S\,IV-Mn\,III 1073\,\AA.~ 
The dominant Lyman $\beta$ line is also indicated.
}
\label{5}
\end{figure}

\begin{figure}
\centering
\scalebox{1}[1]{\includegraphics[angle=90,width=8cm]{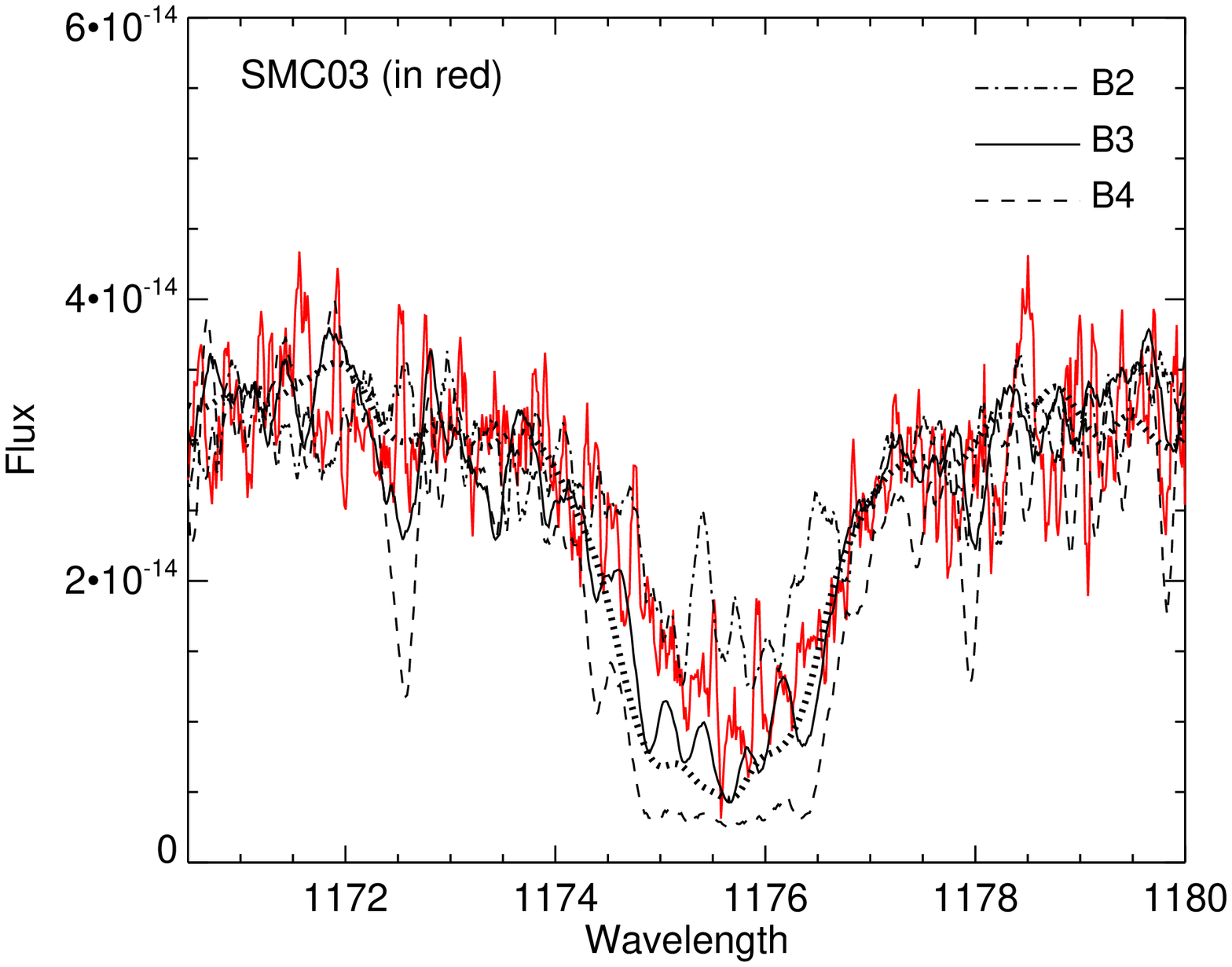}}
\scalebox{1}[1]{\includegraphics[angle=90,width=8cm]{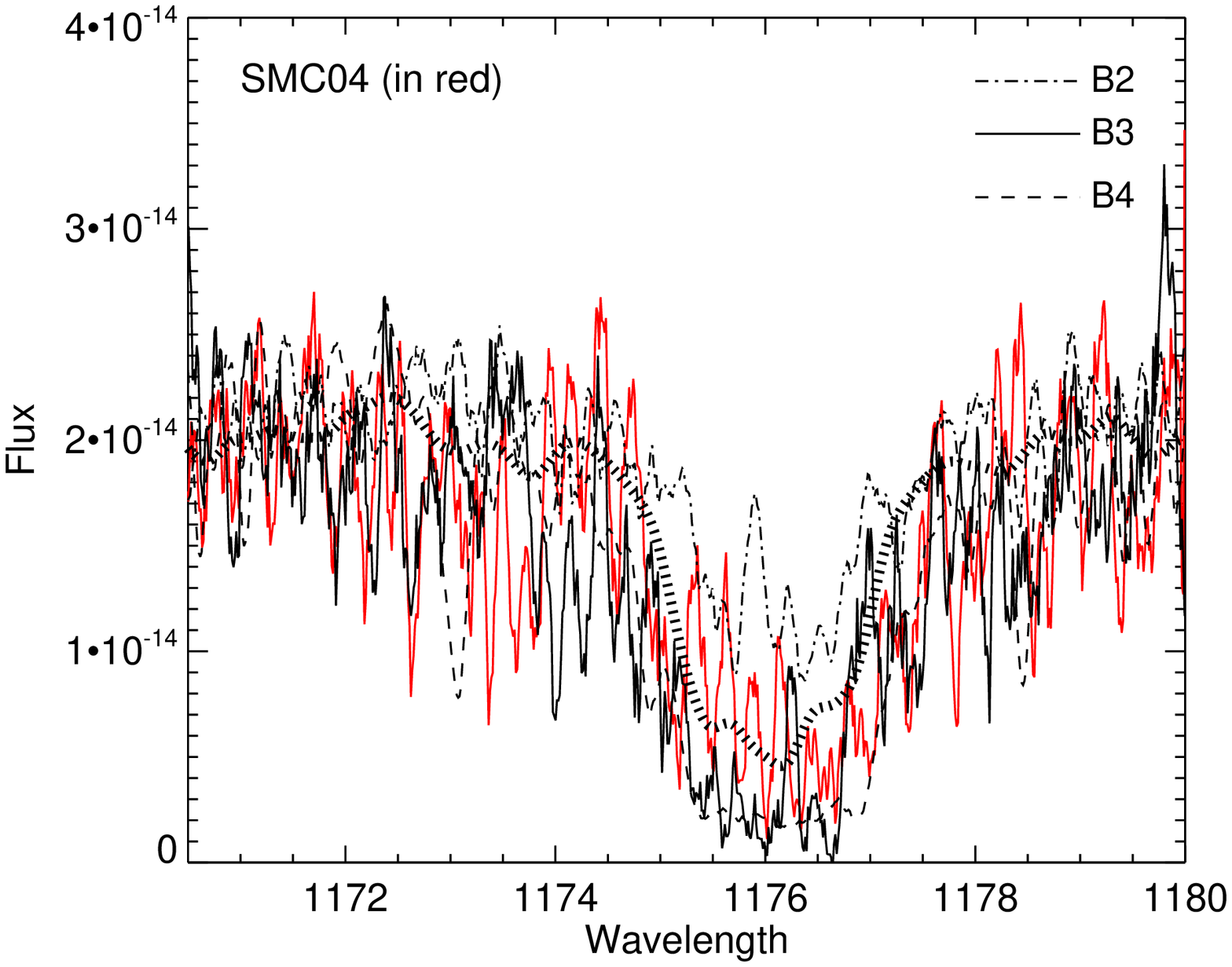}}
\caption{
{\it FUSE}) 
spectra for the region surrounding the C\,III 1176\,\AA~ 
feature in our two program stars, represented by the red line,
and in three Galactic stars representing spectral types B2, B3, and B4 
near the main sequence. For the panel a) these stars are HD\,37367, 
HD\,45057, and HD\,201836.  For panel b) we have substituted the 
spectrum of the B3 star NGC\,330-B30 for the spectrum of HD\,45057. 
Both B3 star spectra give much the same comparison. The dotted line
is the {\sc SYNSPEC} spectral simulation for a T$_{\rm eff}$ = 18\,000K, 
log\,g = 3.5, [Fe/h] = -0.7 model. 
}
\label{6}
\end{figure}

\subsection{Spectral types in the far-UV}


Because the program {\it FUSE} spectra were obtained through the instrument's
Large Science Aperture, we were able to estimate the far-UV flux in addition
to surveying the spectral line strengths.  The spectral lines are consistent
with an early to mid B type spectrum and inconsistent with predictions for
an A star.  We concluded from this that the far-UV flux is emitted by a
secondary star that is too faint to show visible lines in the optical range.
Our initial reconnaissance of the hydrogen Lyman and the metallic lines 
suggested that the spectrum is representative of a star of type B0 to B5. 
To compare our far-UV spectra with those of other metal-poor early B
stars fainter than supergiants, we canvassed the entire {\it FUSE} archive. 
We found only four SMC stars within a factor of ten of the same far-UV
brightness and with O9-B5 spectral types quoted in the literature
(M02, Evans et al.  2006, Blair et al. 2009). 
One of these stars is in the SMC cluster NGC\,330 and is listed by Evans 
et al. as B30. The second star, HV\,1620, is a well known O9 eclipsing
binary.  Two stars are SMC OGLE survey stars that we selected in 2006 for
{\it FUSE} observations. An analysis of these OGLE stars has
not been published and we note specifically that the spectral type given
in the {\it FUSE} archives as ``B2\,V" for OGLE005745.25-723532 was not
based on a spectrum and is not to be trusted. The optical spectrum of
OGLE005100.18-725303 was discussed and given in M06.
Table\,4 gives the coordinates, m$_v$, and spectral types where possible
for these stars.  Of these four stars with known far-UV fluxes only the 
flux of HV\,1620 is comparable to the fluxes of our program stars. 
However, this is a late O star 
and is clearly not a good match to the far-UV spectra of our program stars.
The other three are some 8-10 times brighter, suggesting that they have 
either earlier types, higher luminosity classes or both. The Ly\,$\beta$ lines
and other metallic features in the spectra of the OGLE stars and HV\,1620 
are weaker than those of our 
program stars, suggesting that their spectra lie in the O9-B1 range.

\begin{table*}
\centering
 \caption{Candidate early-type B stars with FUSE spectra }
 \begin{tabular}{@{}lcccl@{}}  
 Object & RA  & Dec & m$_v$ & Sp. Type \\
\hline
NGC330-B30   &  00~56-09.4 & -72~27~58.9  & 14.22 & B3\,III\\
HV\,1620  &  00~54~38.6 & -72~30~04.2  & 14.08 & O9\,V~+O9\,III\\
OGLE005100.15-725303 &  00~58~00.1 & -72~53~03.9  & 13.56 & B1\,II-IIIe  \\
OGLE005745.25-723532 &  00~57~45.2 & -72~35~32.0  & 13.82 & --  \\
 \hline
\end{tabular}
\end{table*}


  In Figure\,5 we show a comparison of the spectra over the range
1000-1070\,\AA~ for each of the two program stars and NGC\,330-B30. 
Although these spectra are offset for clarity, overplotting them 
against one another shows that the wings of the dominant Ly\,$\beta$ 
line are a close match, suggesting that this star is close to B3 in type.
We depict also identifications for 13 available photospheric lines, as
taken from the Far-UV Spectral Atlas of B near the Main Sequence (Smith 2010).
A comparison of the metallic line ratios of the Si\,IV and S\,III relative 
to the Fe\,III and Cr\,III lines, and in turn to the C\,II 1037\,\AA~ line
also suggests nearly equal spectral types for B30 and both program objects.
In addition, this is also true of the possible luminosity class diagnostic
Si\,III~1113\,\AA/Si\,IV~1066\,\AA~ (Pellerin et al. 2002, Smith 2010). 
The value of this ratio is $\gtrsim$ 1 and by itself already suggests a 
luminosity class in the range III-V.  We note again that according 
to our optical spectra the contribution of the secondary star's flux
must be at least three magnitudes fainter than the primary in the
visible wavelength region. This means
that the B secondary has $B$ and $V$ magnitudes of about 17. This is
consistent with the stars' positions on the upper edge of the main sequence,
or approximately luminosity class III and a log\,g of 3.5 to 4.  Finally,
we point out that the far-UV flux of \va~ is some 0.57 magnitudes brighter
than \vb,~ which is probably due to a combination of brighter luminosity 
and/or slightly earlier type.

  With new luminosity class estimates determined, we repeated the spectral
synthesis on the C\,III aggregate at 1176\,\AA.~ According to the
Pellerin et al. (2002) and Smith (2010) spectral atlases of B stars, this 
feature undergoes a maximum for B0-B2\,V spectra. Spectra of stars just hotter 
and cooler than this maximum show telltale secondary lines in the wings of
this aggregate that allow one to distinguish between an O9 and a middle-B 
spectral type.  We found in our spectra that despite the stars' low 
metallicities these features already have strengths close to the maximum 
exhibited in Galactic spectral standards. In addition, owing to the decrease 
in Stark broadening, the C\,III components begin to resolve even for moderate 
rotations.  However, the components are not resolved in our spectra. 
These considerations suggest that the spectral types are in the range B0-B4.
This is in accord with the comparison of the various indicators
just discussed.  In Figure\,6 we exhibit spectra in the region of
the C\,III aggregate compared to B2, B3, and B4
luminosity class V standards. To provide some redundancy we show in Fig.\,6b 
the spectrum of NGC330-B30 as a comparison to \vb~ instead of the Galactic B3
star (Skiff 2009) HD\,45047. In both cases the B3 spectrum shows the best match.
We also compute and display in this figure 
the spectrum of the C\,III aggregate for a representative
photospheric model (T$_{\rm eff}$ = 18000\,K; log\,g = 3.5, [Fe]=-0.7). 
The fit is also good.


If we regard  NGC\,330-B30 and OGLE005100.18-725303.9 as secondary 
spectral standards for a low-metallicity early-type B star, the 
similarity of their profiles with the profiles of our program stars 
indicate that the profiles of the latter are consistent with isolated 
stars, although there is a hint of weakening in the C\,III feature of \va.~
 This inference is confirmed by the agreement of the synthesized C\,III
profile with the \vb~ feature, with again a hint of weakening of a several
percent in the case of \va.~ The depths of these lines suggest that
the contribution of the A primary can be no more than a few percent.
We consider it highly unlikely that our targets are interlopers 
happening to be visible in the {\it FUSE} aperture. 
Rather, it appears that the B stars are secondaries with bright giant or
supergiant A companions and that they are representatives of binaries 
caught at similar evolutionary stages.  Our discussion of radial 
velocity variations to follow confirms this inference for ~\vb. 
In this picture the B and A components should have approximately the same
stellar mass.
In addition, in either of the two wavelength regions where the spectrum 
of one or the other binary component dominates, we cannot detect the
flux from the other.

According to the model atmospheres we used, a B3\,V star's flux (assumed to 
be T$_{\rm eff}$ = 18\,000\,K and log\,g = 4) at a wavelength of 3933\,\AA~ 
should be five times the flux of an A5 (8\,500\,K) of the same radius.  
Recall that we noted above that we should be able to barely discern
a hypothetical 6\% contamination to the K line from the B star. 
This means that the A star has to have a radius of some five times the
B star's for the B star not to dilute the K line flux. This is consistent
with the stellar radii of middle A supergiants provided that the B star has
a radius of $\sim$7R$_{\odot}$. This gives a radius of $\sim$30R$_\odot$
for the A supergiant. This is consistent with values in the literature (e.g.,
Verdugo et al. 1999) and the mass estimated from evolutionary tracks of
$\approx$ 9 M$_{\odot}$ in M10,  and for simplicity we will take 
9 M$_{\odot}$ for the secondary mass too. Based on our models, even with its
larger radius the A primary should not contribute to the flux at 1176\,\AA.~ 
We conclude from this analysis that each component of these binaries
overwhelms the flux of the other in its dominant wavelength regime.
The primary conflicting feature with this overall
description, as already noted, is the strong H$_{\alpha}$ emissions
in both optical spectra, which is 
usually an attribute of a Be rather than an Ae star.


\subsection{Radial velocities}

\subsubsection{Optical lines}
\label{optrv}

Before conducting a quantitative analysis of the spectral lines in 
the disk, it was necessary to first identify them and then determine their 
radial velocities.  To obtain radial velocities from the optical spectra, we 
correlated measured wavelengths of identified lines with their theoretical
wavelengths (Kurucz 1993). 
The results are summarized in Table\,5.  For \vb\ we give the mean of the
``main" (red) component (which is taken as the velocity of the A primary),  
as well as the velocity of the blue component.  To verify the wavelength
systems from our Th-He-Ar comparison spectra, we utilized several telluric
lines from the oxygen B-band (absorption) and the Galactic ISM component in 
the Na\,I D lines.  The internal errors, measured 
by weighting the internal {\it rms}'s for individual ions, 
do not exceed $\pm$4.4 km\,s$^{-1}$, and so we quote $\pm$5 km\,s$^{-1}$.
\va\ shows a weighted average velocity of 105, 106 and 108 km/s
in the 2002, 2007 and 2009 spectra, 
i.e., it shows virtually no radial velocity (RV) variability. 
The spectra of \vb\ disclose mean RVs of of +100, +108 and +161 km/s 
at these epochs.  Although the RVs are clearly variable, we will 
exercise considerable caution in interpreting these differences below.


 We noticed moderately strong $V$ and $R$ emission components
in the lower members of both the Balmer and Paschen series.
The relative strengths of these features decrease to invisibility
at H$_{\zeta}$. These lines are flanked by absorption wings. Typically
the blue peak is stronger, and they have a deep central absorption.
As Table\,6 shows, the
central absorption cores are found blueward of the centroid of the $V$
and $R$ emission peaks, typically by 4-8 km\,s$^{-1}$.
This table displays small differential shifts in these cores as one
progresses up the Balmer sequence from H$_{\beta}$ and a smaller 
peak separation at lower order Balmer lines. The Balmer emission decrement 
is steep, meaning that emissions quickly drop off from H$_{\alpha}$ toward 
the intermediate Balmer lines. 
All the above are signatures of  quasi-Keplerian optically thin 
H$_{\alpha}$-emitting disks.
 
In  most metallic lines of \vb, remarkably, we also found sharp blue
absorption components (discussed later as ``BACs"). These have a blue 
shift with respect to the primary absorption line components. These shifts 
average -48,  -50 and -36 km\,s$^{-1}$ for the 2002, 2007 and 2009 spectra,
respectively. We noticed that in 2009 the BAC is stronger than the
main component for the Na\,D and Fe\,II group lines.


\begin{table*}
\centering
 \caption{Summary of heliocentric radial velocities (in km/s). For \vb~ 
we give the velocity of the main component and/or the associated BAC. 
The number of lines included in the averages is also listed.
Errors reflect the $rms$ of the RVs per line within an ion. 
For H\,I we consider central absorptions in the higher level Balmer
lines.  The mean RVs exclude He\,I lines. }
 \begin{tabular}  {@{}ccccccc@{}}
  \hline
Ion & \va~&\va~ & \va~&\vb~&\vb~ &\vb~\\
\hline
 & 2002 & 2007 & 2009 &2002 &2007 &2009 \\
\hline
CaI   &- &108$\pm$12 (2) &- & -& -&- \\
CaII  & - &106$\pm$5 (5)& 111 (1)   &- &53$\pm$7 (2); 109$\pm$1 (2)&- \\
CrII   & 106$\pm$8 (5)  & 107$\pm$11 (10) & 105$\pm$6 (8) &48$\pm$10 (3) &55$\pm$11 (6);101$\pm$10 (3)   & 125$\pm$9 (5); 164$\pm$9 (8) \\
FeI    &105$\pm$6 (11)   &110$\pm$12 (27)  & 116$\pm$10 (4) &   52$\pm$5 (10); 97$\pm$3 (3)&50$\pm$3 (2) &    168$\pm$5 (11) \\
FeII   &104$\pm$10 (22) &106$\pm$3 (41)& 109$\pm$3 (37) &51$\pm$9 (16); 103$\pm$4 (11)   &54$\pm$5 (21); 105$\pm$6 (30) &119$\pm$3 (14); 167$\pm$7 (23)\\
HI      & 105$\pm$3 (18) &106$\pm$3 (28) &- &106$\pm$3 (16) &106$\pm$6 (28)    &-\\
HeI   &37$\pm$0 (2)  &- & 27$\pm$3 (2)&- &-&- \\
MgI   &105$\pm$2 (3) & 103$\pm$6 (3)  &110$\pm$1 (2)  &54$\pm$4 (3); 97$\pm$4 (3)&58$\pm$1 (2); 104$\pm$3 (2)&130$\pm$1 (2); 162$\pm$2 (2) \\
MgII  &- &115$\pm$24 (3) &- &-&48$\pm$16 (4); 94$\pm$4 (3)&156$\pm$ 2 (3); 175 (1) \\
ScII   & - &109$\pm$5 (2) & 113$\pm$1 (2) & 52 (1)& 69$\pm$7 (2)& -\\
SiII    & 105$\pm$2 (4)   &106$\pm$27 (7)   & 116$\pm$9 (4) &55$\pm$4  (2)&67$\pm$10 (5); 119$\pm$13 (2)& 169$\pm$7 (5)\\
SrII   &  - &106$\pm$13 (2)&-   &-&60$\pm$2 (2)&  172$\pm$5 (2)\\
TiI     & - &98$\pm$26 (7) &107$\pm$5 (3)  &-&48$\pm$12 (7)&   164$\pm$10 (5) \\
TiII    & 106$\pm$2 (7) &105$\pm$7 (35) &108$\pm$4 	 (22) &50$\pm$5 (10); 96$\pm$5 (5) &52$\pm$7 (6); 111$\pm$16 (9) &129$\pm$5 (14); 170$\pm$15 (12)\\
\hline
mean & 105 $\pm$ 1 &107$\pm$4 &111$\pm$4 &52$\pm$2; 100$\pm$4 &56$\pm$7; 106$\pm$7 &126 $\pm$ 5; 166 $\pm$ 5 \\
w. mean & 105& 106 & 108 & 53; 100 &58; 108 & 129; 161 \\
\hline
\end{tabular}
\end{table*}

\begin{table*}
\centering
 \caption{Heliocentric radial velocities (in km/s)  for H\,I emission line components and shift of the central absorption relative to the centroid of the red and violet emission peaks for years 2002/2007/2009.
NP means not present. Note the larger peak separation ($\Delta \lambda$) in higher order lines.}
 \begin{tabular}{@{}ccccccc@{}}
  \hline
Emission line &star & blue-peak &central-abs &red-peak&shift&$\Delta \lambda$ (km/s) \\
\hline
H$_{\delta}$ &\va     &       -/20/NP&     -/107/106&   -/172/NP&-/-2/-   &-/152/-\\
H$_{\gamma}$&\va&    -/31/21&     -/93/92&   -/168/190&-/-6/-14     &-/137/169\\
H$_{\beta}$&\va   &           24/32/27&     95/92/94&   159/166/160&-4/-7/1&135/134/133 \\
H$_{\alpha}$&\va &       55/46/41 &    93/91/91&   153/152/142&-11/-8/-1&98/106/101\\
H$_{\delta}$ &\vb&        -/NP/NP&     -/104/140&   -/NP/NP&-/-/-&-/-/- \\
H$_{\gamma}$&\vb&    -/NP/48&     -/106/110&   -/NP/NP&-/-/-&-/-/- \\
H$_{\beta}$&\vb&          46/NP/-&    105/112/101 &183/181/185      &-16/-/-&137/-/-\\
H$_{\alpha}$&\vb&        55/64/51&     114/111/106&   171/165/170&-1/-4/-4&116/101/119 \\
\hline
\end{tabular}
\end{table*}

\subsubsection{Far-UV lines}

   We used the Far-UV Spectral Atlas of B Stars Near the Main
Sequence (Smith 2010) to identify a number of lines in our {\it FUSE} 
spectra to measure radial
velocities of the B stars at the time of our observations. The
results for SMC-SC3 and SMC-SC4 are +138 ${\pm 12}$ km\,s$^{-1}$ and
+136 ${\pm 15}$ km\,s$^{-1}$, respectively. {\it FUSE} spectra are
prone to several systematic sources of errors in their wavelength
calibrations (Dixon et al. 2007), and thus the zeropoint 
differences may easily
be as large as ${\pm 20}$ km\,s$^{-1}$. Nonetheless, we noticed that
the measured wavelengths of the same lines for the same spectra closely
coincided in nearly all cases. On the basis of the spatial positions of the 
two dimensional spectra, we believe the apertures were placed consistently
over the stellar images and that the radial velocity
difference between the B star spectra is no more than 10  km\,s$^{-1}$.

\subsection{Spectral synthesis methodology}

  Much of the analysis in this paper relies on spectral line synthesis,
so we first describe the methods used for this analysis.  We have 
adopted the results of the fitting of the wings of the Ca\,II K-line 
and assumed stellar parameters for \va~ and ~\vb, namely
log\,g = 2.5 and 3, which are appropriate for luminosity class I and II
mid-A stars, respectively,
and a metallicity of 0.2$\times$ solar
(e.g., Dolphin et al. 2001, Mighell et al. 1998, Dufton et al.
2005).  The spectra were analyzed using the {\sc synspec} and {\sc
circus} (disk) spectral synthesis programs (Hubeny, Lanz, \& Jeffery 1994,
Hubeny \& Heap 1996).  

Using {\sc circus}, we
were able to estimate rotational broadenings, V\,sin\,$i$, from the
absorption lines in both the {\it FUSE} and optical spectra of our
two program stars. For SMC-SC3 the A primary and B secondary have
broadenings of 20 ${\pm 5}$ km\,s$^{-1}$ and 50 ${\pm 10}$ km\,s$^{-1}$.
For SMC-SC4 the corresponding rotational broadenings for the A and B binary
components are 28 ${\pm 5}$ km\,s$^{-1}$ and 75 ${\pm 10}$ km\,s$^{-1}$.

For {\sc circus} analysis one must assume an input
temperature for a putative intervening disk structure, $T_{disk}$, and
several other parameters.  The {\sc circus} program permits as many as
three separate absorbing or emitting regions to be specified, each with its 
own areal coverage, local temperature, and radial velocity. We utilized 
this feature in some cases to specify two distinct temperature regions, 
since the line identifications indicated that the lines we modeled arise
in regions having a temperature range of 5-8\,kK. At these temperatures 
the fraction of the ions considered plateaus, and thus the errors in this 
parameter are not large.
The strengths of disk components of the Na\,D and K\,I profiles 
can be most easily modeled by temperatures of 5\,kK or less, 
consistent with the picture from the H$_{\alpha}$ emission profiles of
the disks extending out to at least several stellar radii. Nonetheless,
our models show that the column densities required to fit these
components decrease monotonically with temperature. Thus these values
are not well constrained. 
The strengths and widths of the Fe\,II and Si\,II lines 
suggest a microturbulence $\xi$ of about 10 km\,s$^{-1},$
and we assumed the disk to be approximately Keplerian. For lines having
both photospheric and circumstellar components, these components 
exhibit no net shift. Therefore, the latter structure appears to be
in a Keplerian orbit. We have
also assumed that the foreground disk segment fully covers the star.  
However, if our lines of sight toward different portions 
of the star sample different disk conditions along the way, our 
column densities will be underestimated. 

These assumptions required the matching of the computed equivalent widths to
the observations using a one or two parameter fit in disk and column densities.
The columns needed to fit the observed line strengths and the strength
ratios of lines arising from the same ion indicate that the strong disk 
lines we measured are optically thick.
In some cases the lines are observed to have peak equivalent widths (EWs)
as large as we can compute them even from assumed optimal conditions.
We note also that for the optical 
strong Fe\,II lines, we fit {\it emission} profiles in
our models with temperatures for a medium with a larger
projected area than the star's - that is, as if
they are formed in lines of sight that do not intersect the star disk.
We will express the column densities in examples below by using the 
unit ``stellar area." 

Our modeling represents an oversimplification of the true and unknown
disk conditions because the disk geometry and velocity fields are poorly
known. In addition, the compututation of line strengths requires a 
common estimated excitation and ionization temperature.
Nonetheless, for exploratory purposes we believe that our models provide
insight into the thermal and velocity differentiation within the disks. 
Generally, departures from LTE in the line transitions will result in 
an underpopulation of the higher levels relative to lower and resonance 
levels, and therefore an increase in the absorbing column density 
for lines arising from excited levels. 
We have computed our absorption column densities in this scattering 
approximation in our {\sc CIRCUS} simulations.

\subsection{The disk features in \va}

  In view of the implied complex nature of the structures
surrounding both program stars,
we will discuss quantitative fits to several line profiles mainly to
get an idea of the column densities, radial velocities, and turbulences 
in the disk gas. Except for a model fitting to the excited He\,I line 
and resonance lines (which form in cold media which we cannot specify
accurately), we will limit our quantitative analysis to the optical
metal lines arising from atomic levels of a few eV. 
Because of the uncertainties involved, we discourage the reader from 
not interpreting our numerical results too literally. 

   Nearly all the features in the optical spectra of SMC-SC3 are
characteristic of absorptions in an extensive and flattened,
differentiated disk or envelope.
The high-level Balmer and Paschen lines can be resolved out to H30 (Fig.\,1a) 
and P24, but any disk contribution to them cannot be 
well determined because these features are common in A supergiants
spectra. Our {\sc SYNSPEC} simulations of these high atomic levels 
lead to estimates for the characteristic density in the line formation
in the atmosphere of 1-3$\times$10$^{11}$ cm$^{-3}$. 

  In our initial modeling we discovered that the disk
temperature and therefore the column density through the disk along the
line of sight cannot be constrained to a single temperature, particularly for
metallic absorption lines in the blue/near-UV region. For example, given
assumed disk gas temperatures values of 6\,kK and 5\,kK, we are able to 
generate good {\sc CIRCUS} fits to weak lines in the 3800-3850\,\AA~ region in 
our spectra with column densities of about 3$\times$10$^{22}$ cm$^{-2}$ or
7$\times$10$^{21}$ cm$^{-2}$, respectively. These values carry uncertainties 
of at least a factor of three and assume full disk coverage of the star.
We were able to get a better handle on these
parameters by finding a region of the spectrum at 5250-5280\,\AA~
that contains both Fe\,I and Fe\,II intermediate strength lines.
Our modeling for the region of  5250-5280\,\AA~ is depicted in Figure\,7. 
The ionization temperature in the disk is already well constrained
by the Fe\,I/Fe\,II ratio to be $\approx$7\,kK. However, we found that
to reproduce also the observed strengths of the features with a 
metallicity and column densities obtained for fits of our other 
line requires that lines of each 
ion be formed in media that favor each of their ionization states,
rather than an intermediate value near 7\,kK. We were able to obtain 
reasonable fits for both disk spectra with a two-component model 
having T$_{disk}$ = 6\,kK and 8\,kK. (In reality this means that the disk
is described by a continuous distribution warm to cooler temperatures.)
The column densities given in 
Fig.\,7 are generally a few times 10$^{22}$ cm$^{-2}$, consistent 
with our less refined analyses of the lines in the far blue.

\begin{figure*}
 \centering
\scalebox{1}[1]{\includegraphics[width=13cm,angle=90]{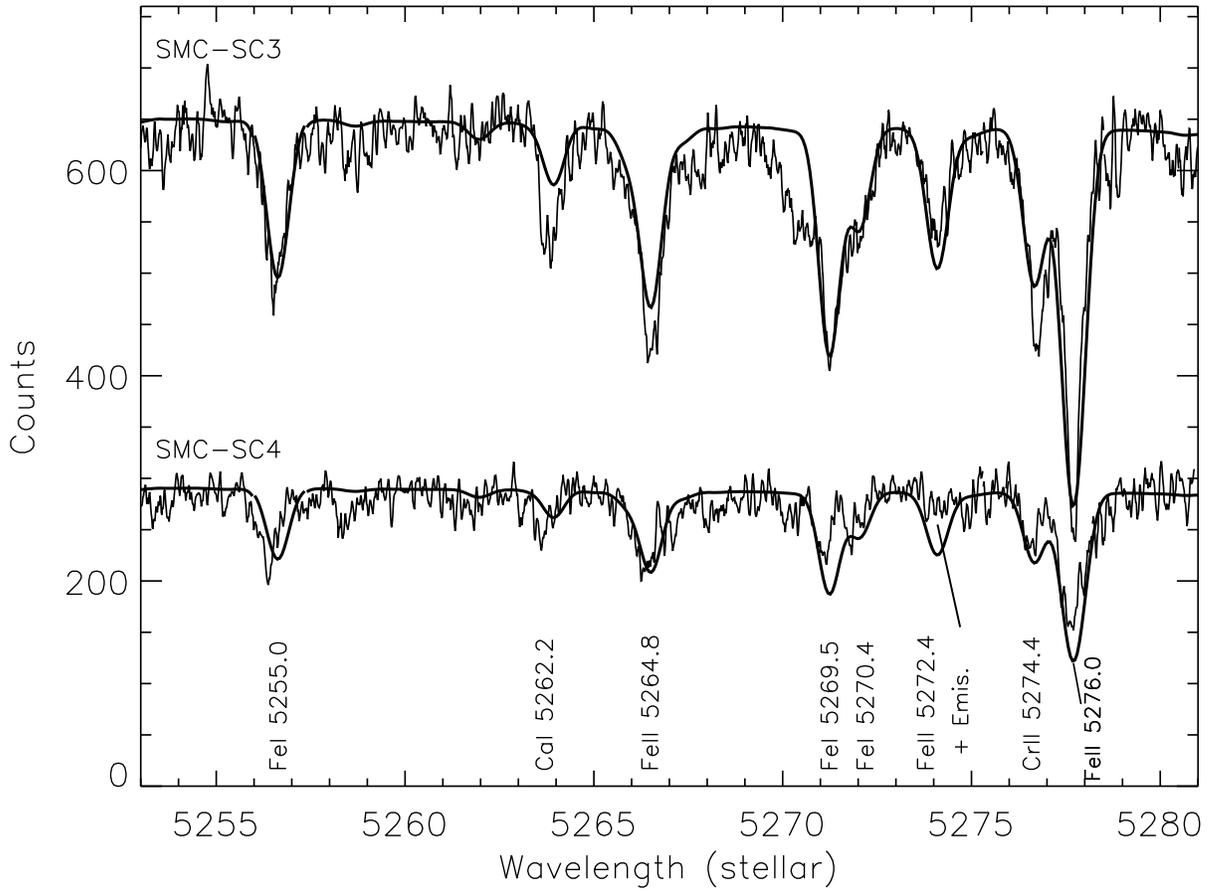}}
\caption{ The fit of yellow Fe-like lines for \va~
(upper) and \vb~ (lower) of our UVES spectra to a two-temperature (8\,kK,
 6\,kK) model.  For the warmer component the column densities 
were 5$\times$10$^{~22}$\,cm$^{-2}$ and  2$\times$10$^{22}$\,cm$^{-2}$ for
SMC-SC3 and SMC-SC4, respectively. The columns for the cool component 
were 3$\times$10$^{22}$\,cm$^{-2}$ and 1$\times$10$^{22}$\,cm$^{-2}$.
The weak feature appearing at redshifted wavelength 5258\,\AA, not present
in our simulation, is likely to be an Fe\,I line the published oscillator 
strength of which is inadequate.  The Fe\,I 5255\,\AA~ line arises from a
10\,eV level. In the SMC-SC4 this feature exhibits weak redshifted emission.
}
\label{7}
\end{figure*}

\begin{figure}
\centering
\scalebox{1}[1]{\includegraphics[angle=90,width=8cm]{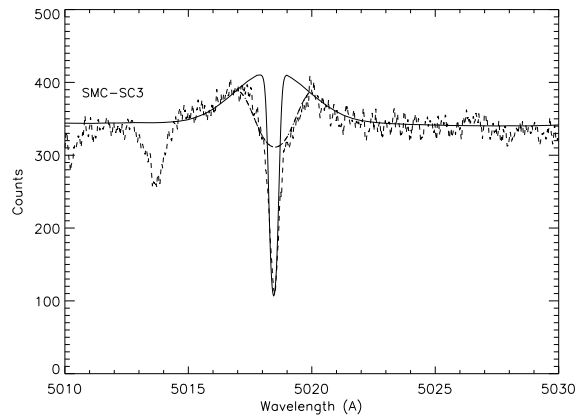}}
\caption{
A fit of the Fe\,II 5018\,\AA~ line (UVES spectrum) for \va. This profile 
was fit by two independent models, one with a temperature of 8\,kK (see
text for other details).
The second component was fit to the emitting wings with the same gas 
temperature and by assuming an area of 11 stellar areas and a gaussian
turbulence function of 70\,km\,s$^{-1}$. Wavelengths are in the observed
system for the 2002 epoch.
}
\label{8}
\end{figure}

A salient attribute of the strongest Fe\,II lines of \va,~ 
(which arise from a common multiplet having 2.9\,eV)  is that
they exhibit broad absorption components and even broader emissions.
Certainly, the emission components and even much of the absorption
does not appear to be photospheric.
For example, among the
strongest three lines, 5169\,\AA, 5018\,\AA, and 4923\,\AA\, the stronger
the line, the more pronounced the absorption and overlying emission wings,
a point which we now elaborate.

   Among the strong disk Fe\,II lines in this spectrum, we chose
to fit the 5018\,\AA\ line, which is depicted in Figure\,8.
Our fit for this feature was achieved by confirming the column density of 
a disk from the emission component of a few Fe\,II lines. These lines are
4923\,\AA, 5018\,\AA, 5169\,\AA, 5316\,\AA.~ We estimated from their
relative emission strengths that that they are mildly optically thick,
$\tau_L$ $\sim$ 3.  For the optical spectrum of SMC-SC3 this corresponds to 
a column density of 3$\times$10$^{22}$ cm$^{-2}$ and a temperature of 8\,kK.
This is the same temperature and nearly the same column density as 
we found for the lines in Fig.\,7. This agreement allowed us to fit
the absorption features of these Fe\,II lines straightforwardly.
The full profile was fit in {\sc CIRCUS} with a two temperature-component 
simulation. We fit the emission using the same
column density and temperature as just found for the absorption component. 
Next we needed to broaden the simulated emission feature by a 
gaussian macroturbulence of 70 km\,s$^{-1}$. The amplitude of the 
emission required adding a free parameter, the emitting gas area, which
we found to be 11 stellar areas in this case.  
We note here that we actually have no independent handle on the
volumetric density of the matter in which the Fe\,II lines are formed. 
Using the column lengths of $\sim$3R$_{*}$, characteristic of the
derived emitting areas ($\sim$11 stellar areas), we can 
estimate characteristic densities of 3-10$\times 10^{9}$ cm$^{-3}$. 
Values of this order of magnitude should not result in visible 
forbidden features, and indeed they are not seen.

  The Si\,II and O\,I triplet in the \va~ spectrum are formed
in plasma with temperatures in the range 6-9\,kK as well.  Although 
their cores are sharp, and consistent with being formed in a 
medium distant from the star, their strong wings suggest a second
component formed closer to the star. 
The Na\,I D and K\,I multiplets in this spectrum have the same 
combination of sharp core and extended wings. This is also true for
several of the Fe\,II lines and the Si\,I 6347\,\AA, 6371\,\AA\ doublet,
which have high optical depths and therefore characteristic mean 
formation sites in the outer disk where the temperature and column
length are low.
For the Na\,I and K\,I lines we could fit the line cores with a column
above the photosphere of 5$\times$10$^{21}$ cm$^{-2}$ for T$_{disk}$ = 
4\,kK and 1.5$\times$10$^{22}$ cm$^{-2}$ for 5\,kK.  

 We address next the remarkable emissions
in the He\,I 5876\,\AA,~ 6678\,\AA,~ and 7265\,\AA~ lines, all of which
arise from 21\,eV levels. We show the stronger of these lines, 5876\,\AA~
and 7265\,\AA,~ in Figure\,9, corrected for their mean redshift 
of 32 km\,s$^{-1}$.
These lines are the only ones in the spectra of both stars that differ
markedly from the other line system(s).  The observed 5876\AA/6678\AA~
line emission ratio is about 3 to 1, which suggests that these lines are
formed in an optically thin medium. Such emissions are typically produced
in a relatively dense plasma by heated gas, such as in shocks, or in
much more energetic astrophysical venues (like winds of Be X-ray binaries) 
by recombination.  Our 
simulations show varying efficiencies of formation in media of 15 to 23\,kK, 
with corresponding emission regions encompassing of 8-12 to 3-4.5 stellar 
areas and column densities of no more than 
1-5$\times$10$^{22}$ cm$^{-2}$ for the two respective gas temperatures.  
If this heated gas undergoes continuous cooling down to 8\,kK, 
it is likely,  given the small volume of formation implied by the 
He\,I emission widths, that the resulting emission generated in the Fe\,II
lines would be hidden by the much broader Fe\,II emission component 
formed over the larger disk volume. 

\begin{figure}
\centering
\scalebox{1}[1]{\includegraphics[angle=90,width=8cm]{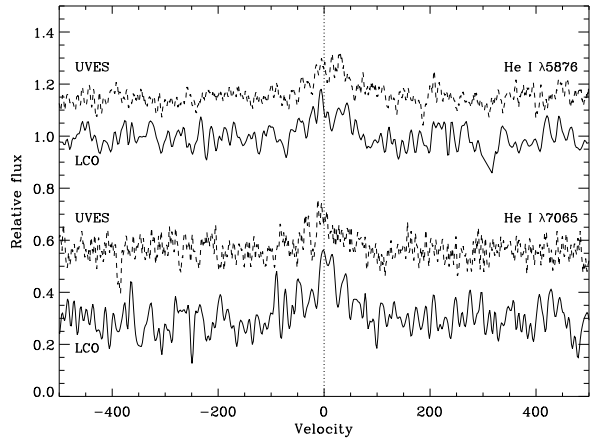}}
\caption{
The regions of the He\,I 5876\,\AA~ and 7065\,\AA~ weak emission lines 
in the spectrum of SMC-SC3 in both UVES (2002) and LCO (2009) spectra.
Here the velocity zeropoint is taken as the mean, -32\,km s$^{-1}$, noted 
in Table\,5 and is -74 km s$^{-1}$ from the average velocity of the 
features in the optical spectrum. 
Note the near absence of velocity shifts between the two. The fluxes
are 4-point smoothing of the raw data.
}
\label{9}
\end{figure}

\vspace*{.15in}

\subsection{The metallic line features in \vb}


   Much of what we found for \va~ is true also for \vb, and we
have already pointed out that the inferred temperature distribution
for the disk around \vb~ is indistinguishable from
the \va~ disk.  For the most part the column densities in our fits
to the main (unshifted) metallic absorption line features run a 
factor of two to three times lower. For example, for a 6\,kK model 
the 3800-3850\,\AA~ lines could be fit with column densities of about 
7$\times$10$^{21}$ cm$^{-2}$ and the Fe\,II lines in the 5250-5280\,\AA~ 
region by some 2$\times$10$^{22}$ cm$^{-2}$.


\subsubsection{Blue Absorption Components (BAC) }

Most of the strong lines in this
spectrum have profiles different from those of \va.~ 
The most noticeable difference is the presence of one and 
sometimes two discrete BACs in the SMC-SC4 spectrum.  
These ``BACs" are so named because their morphology is 
reminiscent of the ``DACs" in resonance lines of radiation-driven winds
of hot stars.  A primary BAC generally occurs at about
-50\,km\,s$^{-1}$ to the blue of a redder or ``main" component,
which we so name because the latter's velocity are coincident with
the RVs of the high level Balmer lines. 
The double entries in SMC-SC4 columns of Table\,5 indicate the
commonness of single BACs throughout the optical spectrum. 

The yellow optical region is rich in intermediate strength Fe\,II 
lines with $\chi$ $\approx$ 3 eV, and these lines show prominent
BAC strengths. For example, the fitting of BACs in the
5018\,\AA~ line in the UVES spectrum 
required column densities of 3-10$\times$10$^{22}$ cm$^{-2}$, 
or several times that of the main (static) components. This is
largely because these components are mildly optically thick. 
The pattern of two BACs is common among the strongest of these 3\,eV 
Fe\,II lines. Figs.\,11a and 11b exhibit double component pattern in the
5018\,\AA\ and 4923\,\AA\ lines, respectively. In both figures a ``secondary" 
BAC shifted by about -100  km\,s$^{-1}$ relative to the main (red) 
component is conspicuous.  The BAC strengths for the Fe\,II lines can 
change over time, and Fig.\,11a shows changes for the 4923\,\AA\ line.
As expected, the strengths of BACs among various lines arising from the
same ion and comparable excitation levels increase with line strength.

BACs are relatively weak in lines arising from near-ground state atomic 
levels in low ionization species such as Mg\,I, Fe\,I, and Ti\,II. 
They are particularly noticeable in lines arising from moderately 
excited states of abundant metallic ions like Fe\,II, Mg\,I, Si\,II 
and even the O\,I 7771-5\,\AA~ triplet (9\,eV) - see Figure 10. 
The excited lines in the N\,I 8703-48 multiplet ($\chi$=10\,eV) 
exhibit a similar behavior as the O\,I triplet.
Many of the strongest lines, including the high order Balmer lines,
and the Na\,I D and K\,I 7699\,\AA~ doublets at first seem to show
a shading toward one wing or the other, depending on the epoch. 
However, as the 2007 MIKE spectrum makes clear in Fig.\,2, this 
tapering is due to the blending of two BACs. In this spectrum the
``tapering" is fully resolved into two sharp components. 

\subsubsection{Ephemeral non-BAC features}

  In addition to BACs, absorptions and emission components appear 
often in the extreme wings of the Fe\,II lines. However,
their behavior is in some respects stranger than the BACs.
Both panels of Fig.\,11 show that the flux in one of the wings
can change from absorption to weak emission among members of the
same multiplet in the same spectrum. The difference in the fluxes in
the red wings of the 4923\,\AA\ and 5018\,\AA\ lines is particularly
surprising because their difference in log\,$gf$ is only about 0.1 dex.
These differences can be seen even more dramatically in Figure\,12a,
again in the same spectrum.  Note in the blue wing the strongest
line in a multiplet, 5169\,\AA, shows absorption. As one proceeds
to the weaker members in the series, e.g., 5316\,\AA,~ the wing exhibits
the {\it strongest} emission. 
   
   The increased activity with time for the ``weak" 5316\,\AA\ line
is exhibited in Figure\,12b.
This figure also shows weak absorption in the red wing, i.e., the opposite
wing in which emission is seen. Other lines show this same behavior
(e.g., Fig.\,11b).

As one proceeds to higher excitation lines the secondary BACs
disappear, and often so do the main components. Figure\,13 shows
vestigial red wing emissions of the excited
Si\,II 6347\,\AA, 6371\,\AA\ doublet similar to those just exhibited
in the Fe\,II 5316\,\AA~ profile in Fig.\,12b.

\begin{figure}
\centering
\scalebox{1}[1]{\includegraphics[angle=90,width=8cm]{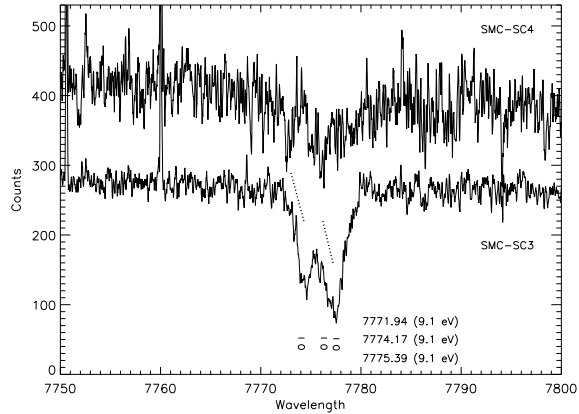}}
\caption{
Comparison of the O\,I triplet in the two program stars (UVES spectra). 
These lines, located at 7771.94\,\AA, 7774.17\,\AA, and 7775.39\,AA,~ arise 
from a level at 9\,.1 eV. Note the broadening and the sharp, blueshifted 
BAC components in the \vb~ spectrum.
}
\label{10}
\end{figure}

\begin{figure}
\centering
\scalebox{1}[1]{\includegraphics[angle=90,width=8cm]{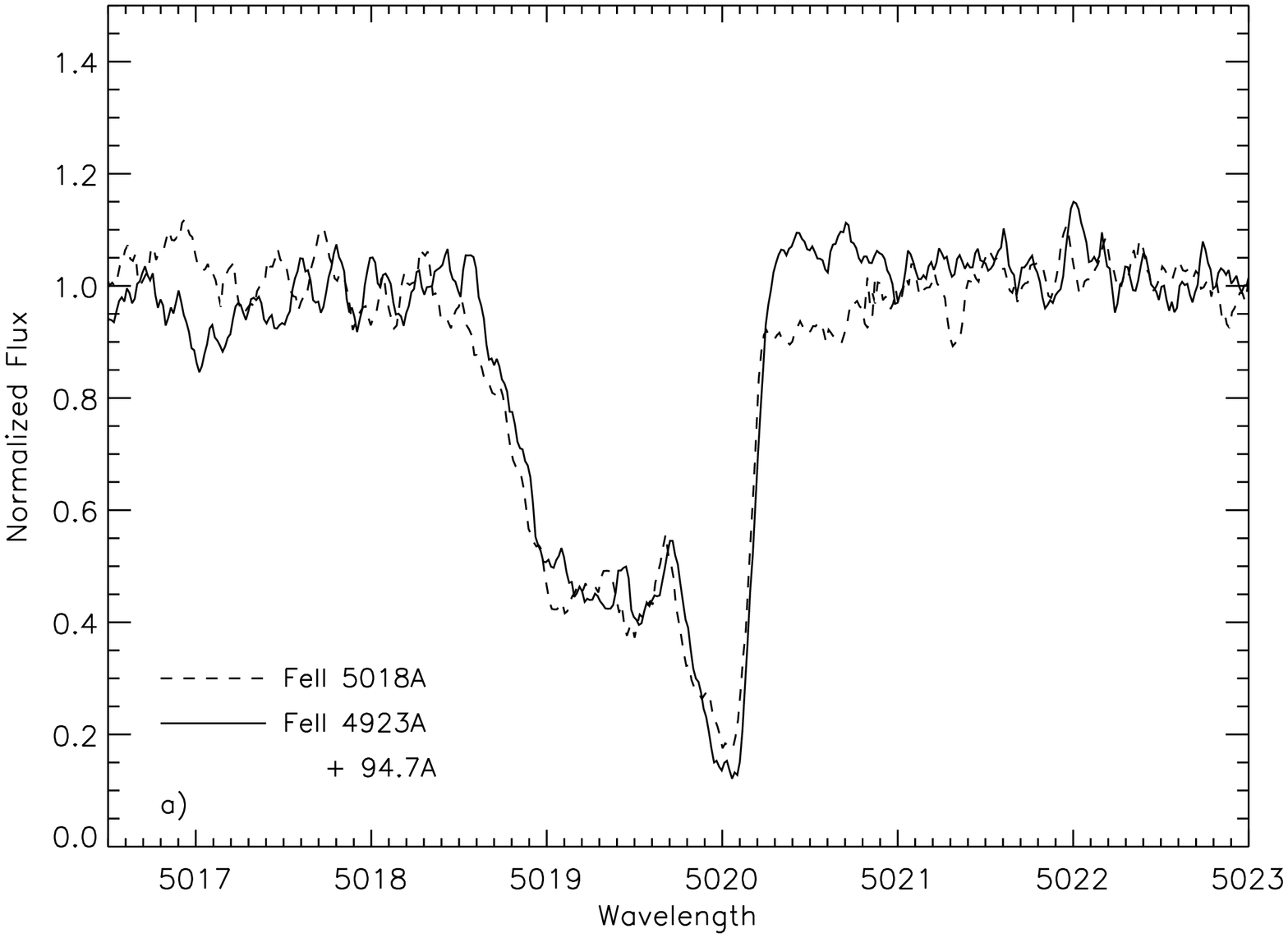}}
\scalebox{1}[1]{\includegraphics[angle=90,width=8cm]{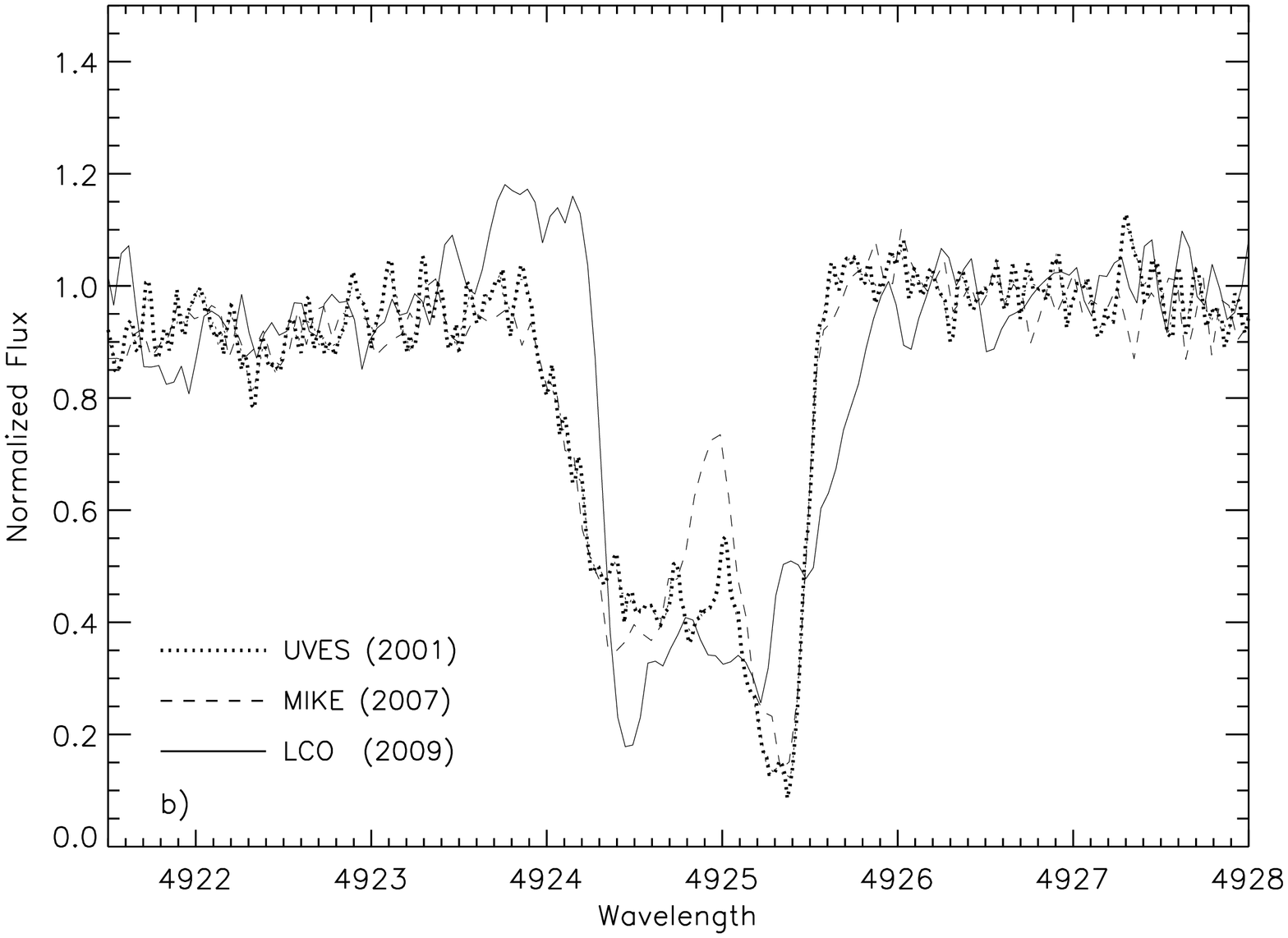}}
\caption{
{\it Panel (a):} 
Overplotting of the Fe\,II 4923\,\AA~ and 5018\,\AA~  lines from the
UVES spectra of \vb, 
exhibiting {\it two} Discrete Blue Absorption Components (``BACs"). The
4923\,\AA~ line was redshifted by 94.7\,\AA.~  The red wing absorption
(5018\,\AA~ line) and emission (4923\,\AA~ line) discussed in the text are 
present in this figure.
MIKE (2007), and LCO (2009) spectra 
of the Fe\,II 4923\,\AA~ profile. The 2007 and 2009 spectra have been
shifted to force the major absorptions to coincide. Red wing absorption
and  blue wing emission 
are present in the 2009 spectrum.
}
\label{11}
\end{figure}

\begin{figure}
\centering
\scalebox{1}[1]{\includegraphics[angle=90,width=8cm]{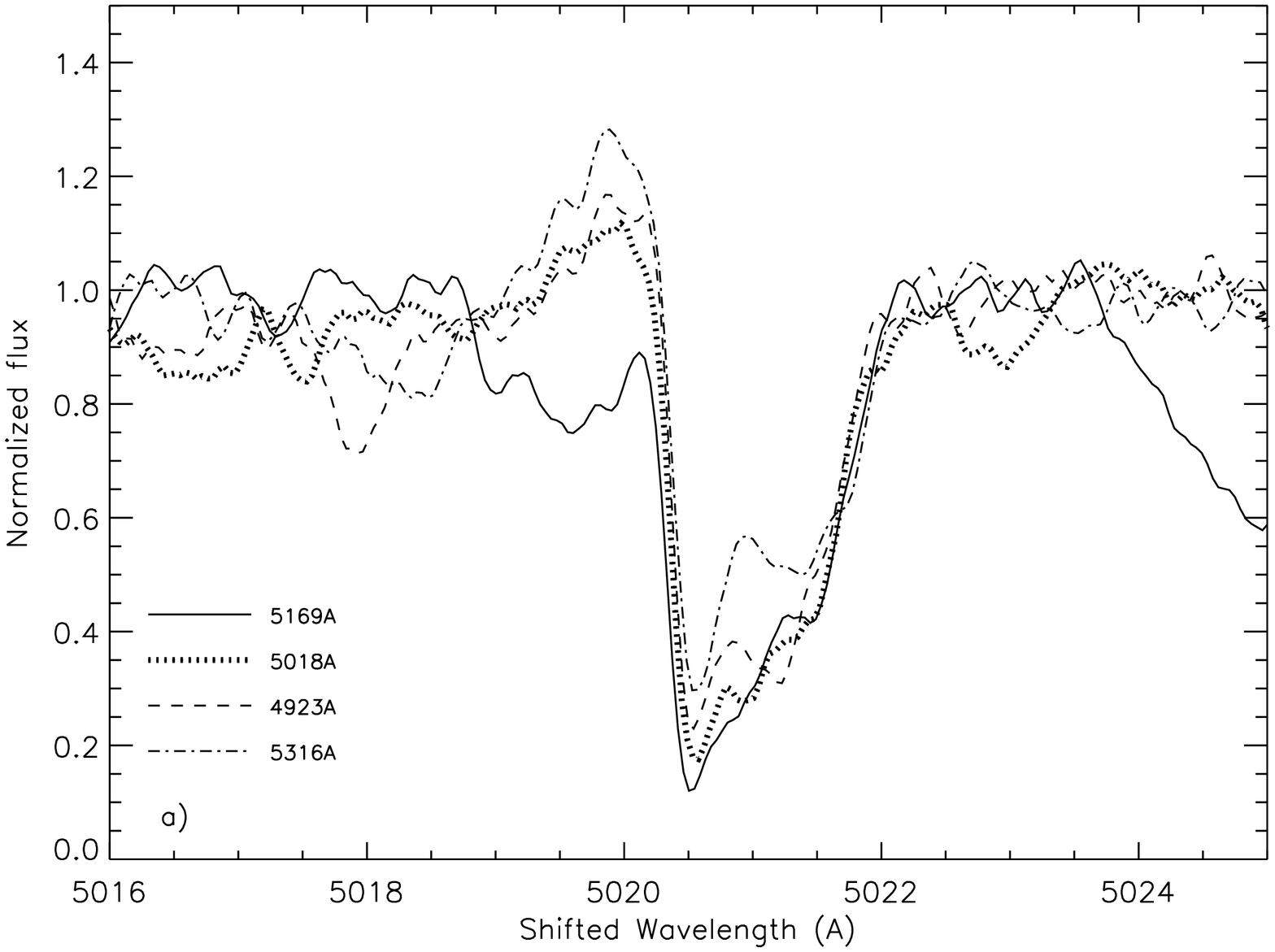}}
\scalebox{1}[1]{\includegraphics[angle=90,width=8cm]{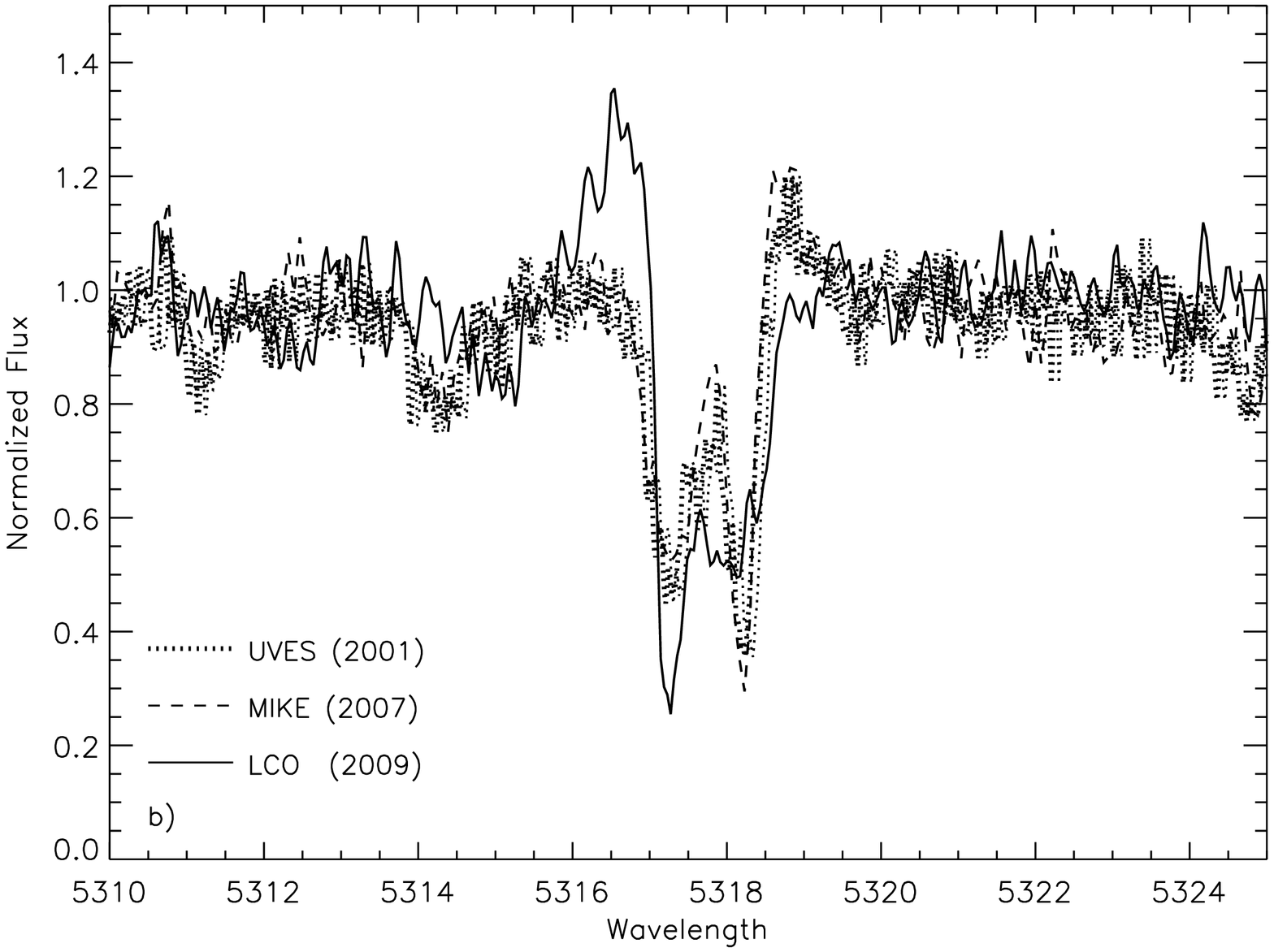}}
\caption{
(a)~ Overplotting of Fe\,II 5169\,AA, 5018\,\AA, 4923\,\AA, and 5316\,\AA~ 
(in order of decreasing strength) in the LCO (2009) spectrum. Each of the
other profiles have been coshifted to the wavelength centroid of 5018\,\AA.~ 
Note the increasing flux in the blue wing at 5019-5020\,\AA) with 
decreasing line strength.
(b) The Fe\,II 5316\,\AA~  line of \vb, exhibiting {\it two} BACs;
the MIKE (2007) and LCO (2009) spectra have been shifted in wavelength 
to force the major absorptions to coincide. The emission activity at 
5316.5\,\AA~ and 5318.5\,\AA~ is discussed in the text.
}
\label{12}
\end{figure}

\begin{figure}
\centering
\scalebox{1}[1]{\includegraphics[angle=90,width=8cm]{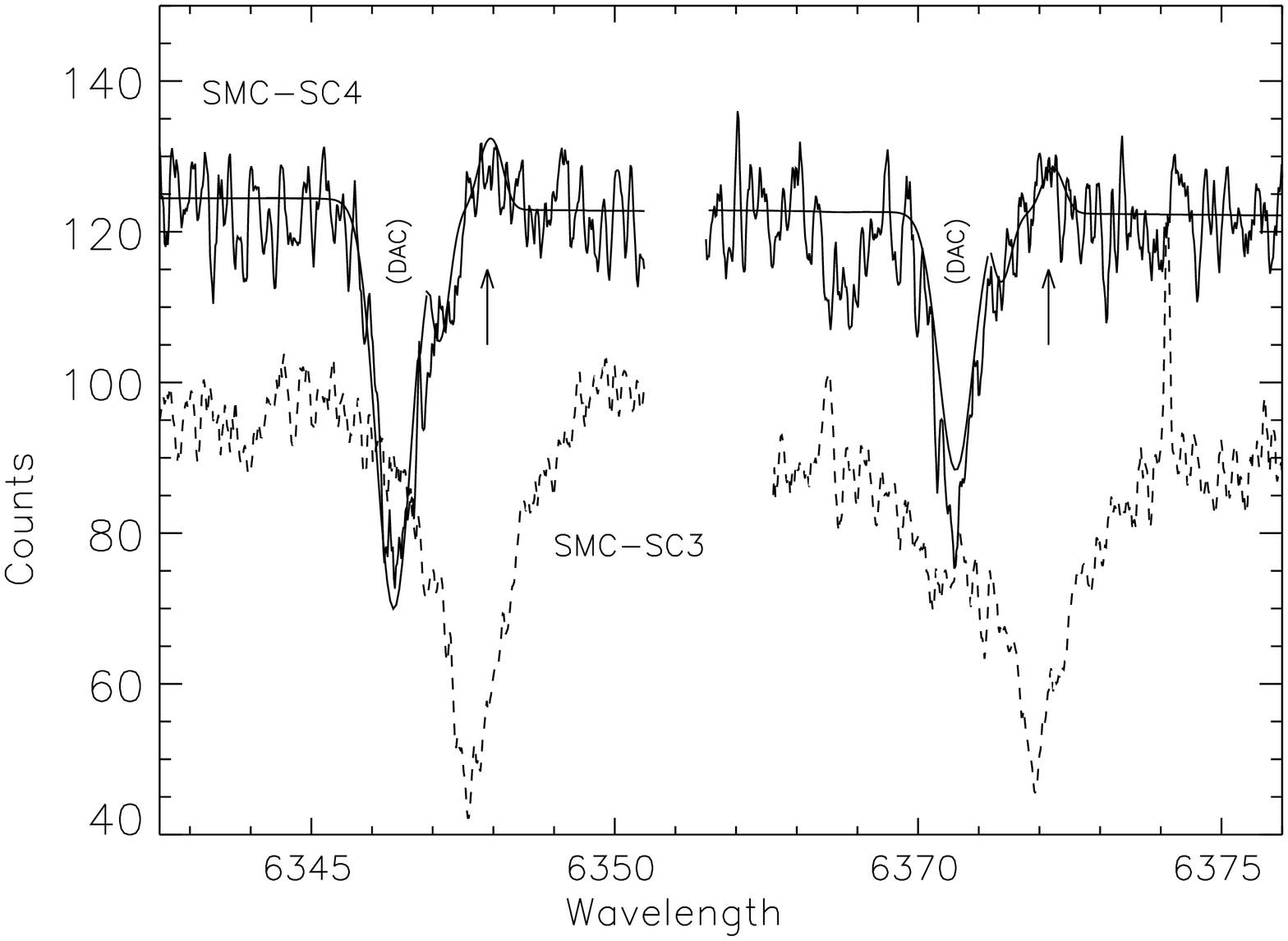}}
\caption{
A fit to the Si\,II  6347\,\AA~ and 6371\,\AA~ doublet (solid line)
of the UVES (2002) spectrum, emphasizing the BACs and the filled in main 
components for the \vb, such as also found in other strong Fe\,II lines
and the excited Fe\,II 5255\,\AA~  line (e.g., Figs. 7, 11 \& 12a).
The observed spectrum is depicted for \va~ in order to guide the eye to
the filled in red wing emission in the \vb~ spectrum. The solid line fit
was attained with a model having a disk temperature of 8\,kK, a column of
3$\times$10$^{22}$ cm$^{-2}$, and a projected surface of 1 stellar area.  
The 6371\,\AA~ profiles have been shifted by -14.9\,\AA~ with respect 
to the  6347\,\AA~ for convenience. 
}
\label{13}
\end{figure}

   In Figure\,12a we point out the curious difference between the BAC
and emission strengths of the same 2009 spectrum: whereas the strengths 
of BACs of lines in the same multiplet or supermultiplet increase as 
expected with oscillator strength, the strengths of the emissions go 
in reverse order, that is, the strongest line ($\lambda$5169) shows 
blue wing {\it absorption} and the weakest line shows the strongest
emission. This is depicted in Fig.\,12a, and the reversal can be seen 
in spectra from other epochs as well. {\it This behavior sets the 
far wing features apart phenomenologically from the BACs.}

   We end this section by reporting that we encountered difficulties 
in attempting to fit the Fe\,II line emissions strengths of Fig.\,12 
with our {\sc CIRCUS} models, even though it was easy to
fit a similar emission in the Si\,II profiles shown in Fig.\,13.
The source of the problem is the combined constraints affecting the
formation of the Fe\,II lines. As one increases the gas temperature
in the models the Fe\,II emissions increase up to a point. However, 
above 9\,kK the flux emitted by a given gas volume decreases rapidly 
as the ionization of iron shifts to Fe$^{++}$. Above this temperature,
regardless of the column density (the lines are optically thick already), 
the emission requires increases in the emitting volume and therefore
the area. In practice, an emitting area of at least 10 stellar areas
is required -- substantially higher values are needed for gas 
temperatures greater than 10\,kK or less than 8\,kK. 
These simulations bring out a contradiction according to our assumptions:
{\it whereas the blue wing absorption in the strong $\lambda$5169 
line can be formed in columns subtending no more than one stellar area,
the emissions of Fe\,II lines in the same multiplet must be produced in
substantially larger areas and volumes and possibly different temperatures.} 
Clearly the formation of various components of the Fe\,II lines in
this spectrum occurs in separate regions having diverse properties. 
We will return to this point in the next section.

\section{Discussion}

\subsection{General}

  The primary attributes that initially drew our attention to the program 
stars were their multiple periods in the OGLE $I$-band light curves. 
However, it was the complex nature of their optical spectra even at moderate
resolution that led to this detailed spectroscopic analysis. We have already 
noted the strong double-peaked emission in H$_{\alpha}$ and other low 
members of the Balmer series, which is uncharacteristic of an A star, 
even one of high luminosity.  The acquisition of far-UV
spectra has clarified that these objects are A + B luminous binaries,
and three optical spectra have confirmed radial velocity variations
for SMC-SC4. The current investigation discloses no new clues
to the origin of the 15 day period of SMC-SC3 or of the ``chaotic" 
photometric activity for SMC-SC4 discussed in M10. However, we can
remark further on spectroscopic activity pertinent to the 184 day 
``eclipse" period.


 In $\S$3.3.1 we expressed doubts as to how far the RV variations 
from the optical lines of SMC-SC4 can be interpreted. For
example, given the extreme changes in line profiles
in the 2009 spectrum, it is not clear that we can trust the 
cross-correlation among spectra of different epochs 
to faithfully represent the motion of the A
star around its orbit.  We also note that in the only blue Balmer line 
covered in all three observations, H$_{\delta}$, the wings suggest that 
the RV variations are larger than 100\,km\,s$^{-1}$. For this reason we do
not trust these few measurements to estimate binary orbit attributes 
such as eccentricity and inclination reliably. 
 Nonetheless, reasonable to infer that the RV variations are
due to the same 184 day eclipse period discussed by M10, even though the
eclipses are almost certainly not due to the secondary star. 
We will comment on further on this geometry in \S4.4.

  If one adopts Kepler's Third Law for the \vb~ orbit
18\,M$_{\odot}$ and a 184 day binary period, one finds
a semimajor axis for the binary systm of 1.66\,AU.
Considering next the disk, if one assume Keplerian orbits and
adopts a measured separation between the $V$ and $R$ emission components 
in the H$_{\alpha}$ profile of 110 km\,s$^{-1}$, 
the characteristic orbital radius for the disk becomes 4.9 sin$^2 i$ 
AU.  Furthermore, the well resolved nature of the two emission peaks 
implies once again that the sin$^{2}i$ factor is not far from unity.
Then assuming further only that the disk and orbital planes are at 
least roughly coincident, the ratio of the size of the disk to distance 
between the stars is about a factor of 2-3 (4.9\,sin$^2\,i$/1.66).
The conclusion is that in fact the emitting gas arises in a 
{\it circumbinary} (CB) disk around the SMC-SC4 system. 
Confirmation of this picture comes from the fact that the V/R emission
ratio of the H$_{\alpha}$ profile of \vb~ changes from the 2002 to the 2007 
epoch even at near-like binary phases. 
We would expect the details of the emission profile to be different
if the emission came from the environment of one of the binary components.
We use the descriptor ``circumbinary" hereafter to describe the structure 
causing the disk emission and most of the exophotospheric absorptions in
its spectrum. We have not made the case for RV variability for SMC-SC3, 
but given the discovery of its B secondary this seems to be a matter 
of discovery given our poor phase sampling.

 As for \va,~ if one again takes
the sum of masses as 18M$_{\odot}$ and a period of 238 days, 
one derives a semimajor axis of 5.4\,AU of the system.
This is larger than the scale of the \vb~ system, but then the H$\alpha$
emission is some four times stronger for \va~ (Table\,3), which implies 
a more extensive disk outside this radius.

   Altogether, the high resolution spectra have revealed a number of novel 
properties that are different for each star and yet reminiscent and of one 
another.  the two stars, and the overarching question is how the evolutionary 
state of the binaries and the remarkably slow rotations of the 
component stars might be responsible. These novel properties include emission, 
the unusual excitations implied by the emissions in He\,I (\va~ only), Si\,II,
and Fe\,II lines, and for \vb~ the presence of multiple components,
including ``BACs" in the strong resonance and intermediate strength 
lines, and emission components in permitted Fe\,II and Si\,II lines, 

 Before evaluating the spectra of of our program stars further, we should 
state our assumptions about the dynamics of the binary-wind system that 
will facilitate consideration of the simplest possible picture that 
at the same time is consistent  with the complex features we see.  
Because of their presume relatively high space density in the region 
of the SMC first investigated, we assume that both program stars 
are immediately post-main sequence. We have also assumed that M$_2$ 
$\approx$ M$_1$, and this is based on the absence of any evidence of
extensive mass transfer, anomalous abundances, or rotational
spin up.  The stellar radii we determined particularly from the surface
gravities inferred from the Ca\,II and C\,III lines suggests that 
the stars' radii are much smaller than their Roche lobes.
This leads to the simplifying picture that, except for wind-wind
or wind-CB disk interactions, the wind kinematics and geometries
are not differ radically from those of single stars. 
Of course, the fate of wind efflux 
stopped by collisions with the other wind or a disk, as suggested by
shock features, suggests a more complex scenario. We assume that the 
residual momentum of the B star wind forces some of this stalled matter 
to the A primary, perhaps in a relatively narrow neck defined by the 
inner Lagrangian point.  If any matter also returns to the B star
it would not be observed in our optical spectra.

\subsection{The emission features in the \va~ spectrum} 

   The H$_{\alpha}$ emissions in both program stars are strong but not
at the high end of the range among classical and pre-main sequence Be stars.
The presence of H$_{\alpha}$ emission has long been used to confine the 
cool limit ``Be phenomenon" in the HR Diagram. Jaschek et al. (1988)
have surveyed a large number of Galactic Ae stars, and even the class
most resembling our program stars in having a rich optical shell spectrum,
their ``Group II," is confined to types A0-A1. The emission profiles of
those early A supergiants are weak, transient, and have a P Cygni, or 
inverse P Cygni, shape (e.g., Verdugo 2005), characteristic of activity 
close to the stars' surfaces. For our program objects the continuum fluxes 
of the B secondaries are several times smaller in the red than the primaries'
fluxes. It strains credibility to imagine that the A stars or for
that matter the B stars' photospheric UV fluxes can excite such strong 
emission. Emission in the H$_{\alpha}$ line is invariably generated by
recombination in dense circum-stellar/binary environments of hot stars. 
Yet the extended wings of these features hint at the importance of 
electron scattering, and this implies that the disk extends to low 
densities and large radii. 

  The presence of $V$, $R$ emissions in strong Fe\,II lines of SMC-SC3 
is a characteristic in a number of well-known classical Be stars, 
These spectroscopic signatures are formed by scattering in
dense, high optical-depth disks as well.  An important characteristic 
of these emission profiles is that they remain almost symmetric in our
observations, suggesting a stationary disk. However, a small persistent 
excess in the $R$ emission may hint at a few km\,s$^{-1}$ expansion of  
this structure.  Separate from these considerations, the superposition
of broad absorption profiles of strong Fe\,II (e.g., Fig.\,8) and other 
excited lines like the Si\,II doublet suggests the 
presence of a separate absorbing source close to the A star component
of SMC-SC3.

 A clue to understanding the energetics of the disk of SMC-SC3 
is the presence of He\,I lines from the primary's photospheric 
and circumbinary lines. As with the hydrogen emissions, these 
lines cannot be excited by the photospheric radiation field.
A ready alternative is the shocks provided between wind-wind
collisions between the stars.
In the absence of profiles of the
UV resonance lines, this is still a speculative idea. Nonetheless, 
the fact that the velocities of the He\,I lines are so different
from all the absorption line velocities indicates that they are
formed in a different physical region that need not even be 
confined to the disk plane. From our fittings of the strengths
of He\,I ($\S$3.5), we found that for plasma heated to 
18\,kK, a column length of 3$\times$10$^{22}$ cm $^{-1}$ is
typical. Adopting a value of N$_{e}$ $\sim$ 10$^{11}$ cm$^{-3}$, 
typical of CS disks of Be and Ae stars, the column length implies a 
radial extent of 3$\times$10$^{5}$ km ($\sim$2\% 
of the A star's radius) for the formation region; if the temperature 
this extent could be much less. 
This is consistent with shock formation and inconsistent with mechanisms 
that excite helium atoms over a large volume, e.g., irradiation by X-rays.

\subsection{BACs and related features in SMC-SC4's optical spectrum}

The kinematics of the disk surrounding the \vb~ system are quite different 
from the \va. For example, the metallic lines do not show a symmetric two
($V$, $R$) component emission that suggests the presence of a stationary
disk nor that the optical depths of the circumstellar or circumbinary
absorptions have as high optical depths.  In contrast, 
there are the blue and occasionally red absorption components, as
well as emissions in the wings of the main line whose positions
seem to be determined by the orbital phase. 

Although ubiquitous, novel, and likely present at all orbital phases, the 
BAC features are not unique to the \vb~ system. In a little noted discovery, 
Heydari-Malayeri (1990) reported that many of the metallic absorption lines in 
the optical spectrum of the SMC supergiant N82 consist of double components, 
the dominant member of which is blueshifted by about -46 km\,s$^{-1}$.
This is almost the same value found for \vb. The Balmer lines also show
double-peaked emissions intermediate in strength between those
we find SMC-SC3 and SMC-SC4.  In addition, Heydari-Malayeri found that
lines of different ions have various red to blue emission
component ratios. This caused the RVs of members of ions exhibiting these
components to have a large RV scatter.
This author classified N82 as a ``sgB[e]" star largely on the basis of the 
presence of infrared emission and the appearance of [Fe] emission lines.
However, an important supporting justification for the author's classifying 
N82 as Be rather than Ae was the star's strong Balmer line emissions. Because 
we now know that this is not necessarily a {\it sine qua non} criterion for 
Be classification, the argument that this object is a B star is weakened.
The presence of [Fe] emission in the spectrum of N82 but not in our program
star spectra suggests that N82's attributes are phenomenologically distinct. 
 However, its spectrum is similar in including BAC components in 
metallic lines and substantial emission in the lower Balmer lines.

Overall, the striking attributes of the BAC-like components in the \vb~
spectrum are:~ {\it 
a)} their ubiquity among many types of lines, 
{\it b)} their consistent blueshifts of -50 to -32 km\,s$^{-1}$ in at least 
three orbital phases, and {\it c)} their concentration mainly among
lines arising from atomic levels of a few eV.
Points {\it a} and {\it b} argue that the BACs appear the same around the
orbital cycle and are perhaps caused by impacts of an axisymmetric outflow.

  We also found, depending on the epoch, 
that absorptions in either the far blue or red wings of some Fe\,II 
lines show an extraordinary sensitivity to excitation and line strength. 
The flux in emission features exhibits a reversed dependence on the line's 
oscillator strength, with the weaker lines showing the greater emission.
The sensitivity of emissions to line strength fits in with the shock
hypothesis because it implies a rapid increase in local temperature, 
perhaps over only 10$^{4}$ km in the case of the 3\,eV Fe\,II lines, 
in the formation length where the emissions of (weaker) lines are formed. 
The sensitivity to line strength and excitational information also hints 
that the energy input into the putative shocks has a small ``bandwidth,"
that is, if the outflows were less or more energetic then some other
range of excited states would be collisionally populated. 
This is consistent with the wind energetics because winds have a
characteristic velocity when they impact stationary matter at a 
fixed distance. According to these ideas, one expects the excitations
of atoms exhibiting emission to depend on the wind velocities, and
thus to some degree also the orbital separation between the stars.
Perhaps these conditions are ``just right" for SMC-SC3 to excite lines 
at $\approx$20-25\,eV (He\,I lines).


\subsection{Sketch of geometry and kinematics, outstanding questions}

   From the  dissimilar nature of the optical and far-UV 
spectrum (and the RV variations for \vb), we have been obliged to
adopt a binary model involving nearly two equal mass A + B components.
The strong double-peaked emission in the H$_{\alpha}$ line against 
the continuum of an A bright giant or supergiant implies the presence 
of a flattened Keplerian disk.

  The amplitude of the radial velocity variations compared to the 
almost stationary Balmer emission cores
shows that at least in \vb~ the H$_{\alpha}$-emitting structure is a
circumbinary structure. We cannot rule out the possibility of  RV 
variations in \va~ on the basis of their apparent consistency in 
just three observations. 
Indeed, the requirements to produce blueshifted He\,I line emission
suggest unusual energetics  in a region that has a different radial 
velocity than the A star. The
orbital separation of the binary components, estimated above for \vb\
to be $\approx$1.7sin\,$i^{2}$ AU, is about ${\frac 13}$-${\frac 12}$ of
the characteristic CB disk radius. The absence of forbidden
emission or metastable absorption lines argues that most of the disk does not 
have a low density. The optical/IR colors imply that reddening from cool dust 
is negligible.

  We envision that each of the binary component stars has a radiative wind, 
with the A star's 
wind being weaker, slower, and perhaps denser near its surface.\footnote{For
ballpark numbers winds of early B stars may 
have~ \.M $\sim$ 10$^{-9}$--10$^{-8}$ M$_{\odot}$ yr$^{-1}$ 
and a terminal velocity of 1000-2000 km\,s$^{-1}$.  Typical values 
for winds of A supergiants are thought to be $\ltsim$10$^{-7}$ 
M$_{\odot}$ yr$^{-1}$ and 300 km\,s$^{-1}$ (e.g., Verdugo 2002).}
These winds will interact along an annulus centered on the line
connecting the two stars. The shock-heating in this wind will show
a maximum intensity at velocities in between those of the stars,
though not necessarily at the barycenter. This would explain the
displacement of emission of the He\,I lines for \va~ and Fe\,II lines
for \vb.~ The wind continuously replenishes the  CB disk, but 
we have no direct estimate of the relative amounts of disk or 
post-shock matter returning to the two stars or of it escaping through
the outer edge of the CB disk. The symmetric emissions in the Fe\,II 
and other lines in the \va~ spectrum have a separate origin, 
and probably originate in the quasi-stationary CB disk.

   Apart from the 1-2 BAC pattern in the spectrum of ~\vb, far wing 
absorptions appear in the opposite wing from far wing emissions ($\S$3.6). 
Their juxtaposition suggests a common origin for these two non-BAC features.
We suspect that their complex morphology arises from their formations 
along different sight lines to the A star and to flowing CS streams in
the sky plane. This complicated behavior is reminiscent of red- and 
blueshifted components in time-resolved ultraviolet spectra in Algol 
systems. Peters \& Polidan (1984, 1998) have noted the existence of 
flows associated with ``High Temperature Accretion Regions" caused by
streaming of wind efflux from the cool giant star as it expands and 
transits through the binary's inner Lagrangian point. This stream 
collides with a disk around the receiving hot secondary. This is
the reverse of the present context, in which the receiver is the
cooler star. In the present case we envision that
the collision points occur behind the orbiting primary and generate 
shocks. These are manifested in spectra as Doppler shifted emission 
components.  Disk matter cools and eventually falls into the 
cool (A-type) star, rather than only through the inner Lagrangian point. 
The result is a complex array of relatively 
high velocity blue- and redshifted emission and absorption components
that change their morphology around the orbital cycle. 
The high positive velocities compared to the A star in the 2009 spectrum 
(Fig.\,11b, 12a) indeed suggests that matter falls toward the A star 
as seen from its trailing side, as Peters \& Polidan found for the Algols.
One aspect of this picture is that it predicts that the shock
geometry can be complex and not necessarily repeatable from the cycle 
to cycle.  Even at the same time the volumes producing emission
for lines of different ions are not the same.

 It is tempting to speculate that the infalling matter inferred from 
our 2009 spectra of \vb~ is associated with the observed eclipses. 
Actually, any such picture must be complicated because it is the ``wrong" 
(2007) spectrum which nearly coincides with the egress from a continuum 
light eclipse.
This would suggest that the stream spirals nearly a complete
circuit (some 65-90\%) around the A star until it settles onto the star
(the 2007 spectrum was obtained beyond the midpoint and during the egress
of a photometric eclipse; see Table 1 here and Figure\,6 in M10).
Spectroscopically, this may be the cause of the 
second blueshifted absorption component of the strongest lines 
(e.g., dashed line in Fig.\, 2, solid line in Fig.\,11b).
At this binary phase the blue component might actually be a better measure
of the A star's velocity in orbit. The 2009 spectrum was taken at the
opposing phase, an infall phase in this picture, before the matter has
piled up and settled to the surface, At this phase the light curve
registers  only a broad, small-amplitude depression.
We note that our picture has some similarities
with that presented by Soszynski (2007) to explain eclipses in
interacting binary systems composed of red giants and low mass stars
in the Large Magellanic Cloud.

\section{Summary and Conclusions}

Even as luminous variables in the SMC, our two program stars exhibit a
extraordinary set of photometric and spectroscopic properties.
These include a multiperiodic photometric
variability, a $\approx$B3 type spectrum in the FUV, an optical A-type spectrum
with Balmer emission lines and metallic lines from an intervening disk. For
SMC-SC4 the spectral anomalies include the presence of one and sometimes two
Discrete Blue Absorption Components (BACs) in metallic lines and RV variations.
We have remarked that the optical spectrum of the SMC supergiant N82 in the
SMC exhibits BACs and strong H$_{\alpha}$ emission.
However, unlike N82, neither of our stars shows forbidden lines.
Because UV spectra are not available for N82, nor an optical RV campaign
mounted, its suggested binary nature has not been tested.

We argue that our
program stars reflect a comparatively brief stage in the life of intermediate
mass $\sim$ 9 M$_{\sun}$ binary components. With separations of a few AU between
the components, the winds of the two stars can interact and produce a large
array of spectroscopic phenomenology. This phenomenology can include emissions
in lines of intermediate and/or highly excited ions (e.g., He\,I) and
time-variable pattern of sharp absorptions, in addition to the BACs.
It is an open question why as many as two objects have such similarly
unusual properties from a sample of only bright eight Mennickent Type\,3
variables.  In any case we may dub these objects prototypes of a
presumably small group of Magellanic Cloud wind-interacting A + B binaries.

  We speculate that the emission components originate from the 
interaction of that component of the A and B star winds in the
zone near the line that intersects the centers of the two stars.
The double peaked emissions in the Balmer lines
disclose an emitting  Keplerian disc, which from the peak 
separation has a larger extent than the binary separation. We have 
suggested, in part because of the He\,I line emissions in the \va~ 
spectrum, that wind-wind interactions cause localized heating and 
ionizing photons capable of exciting the observed level of H$_{\alpha}$
model. 

The Fe\,II line emission strengths are {\it anticorrelated} with the line's 
oscillator strengths. We have attributed this extraordinary circumstance 
to their formation in a geometrically thin but optically thick and two
layered column, where emission is formed in the deeper hotter shock and
the absorption is just outside this region (though in the foreground of the 
observer's sight line). We have also speculated that wind-wind interactions 
in the zone between the stars collisionally excite He\,I line emission and
also produce Balmer emission in the quasi-Keplerian circumbinary disc. 

  We have surmised that the primary BAC (absorption) feature is caused 
by the violent impact of the expanding A star wind into this CB disk. 
However, although we do not have 
enough observations to characterize the time-dependent secondary 
components well, it appears that the strengths of the emission components 
in either one of the far wings of the 3\,eV Fe\,II lines are physically 
associated with absorptions in the opposite far wing. 

  We have speculated that a rich line profile variability, 
including the so-called secondary BACs and emission components, 
is a consequence of a matter stream originating from the wind of the 
B star and settling on the A star.  This is a reverse Algol scenario
in which matter is not  constrained to flow only through the inner
Lagrangian point. Perhaps this settling
processes is responsible for the eclipses in the continuum light
curve of SMC-SC4.

   A number of observing programs can be suggested to test these ideas
and conceivably offer new alternatives: 

\begin{itemize}

  \item A high-resolution monitoring of these objects should be 
undertaken to construct a radial velocity curve and determine
the periods, separations, and inclinations of these systems.
These same data can be used to trace the velocities of the various
components in the primaries' spectra. For \vb~ the features of interest 
are the BACs and the pairs of absorptions and emissions in the wings 
of the Fe\,II lines. For \va~ it is important to determine whether the
velocity of the He\,I lines is constant or shows a small amplitude
variation around the system's velocity. 

  \item UV observations of
the prominent resonance lines are needed to characterize the wind
strength, velocity, and its behavior through the orbital cycle. 
We expect that these features are diagnostics of the B star's wind. 
It may also be possible to trace velocity changes in signatures of 
the A star's wind, such as an arguably Ca\,II K emission feature 
(Fig. 4b).  In general, we expect the profiles of the UV resonance lines 
to be highly complex and variable.

  \item Any understanding of these particular objects rests on
the expansion and adequate definition of this class, which in term entails 
finding more members. So far our small sample is drawn from a group of
luminous optical/IR variables in the SMC. One prong of this search 
would be to find stars whose spectra exhibit BACs.
To date we have obtained moderate resolution spectra of a
number of luminous B stars in the OGLE surveys of both Clouds.
We find that two LMC stars have properties similar to our two prototype
objects. These are OGLE05141821-6912350 and OGLE00552027-7237101. 
In addition, Mennickent et al. (2010) noted that the object OGLE LMC LPV-41682 
(P = 219.9 days; Soszynski 2010) exhibits modulated eclipses 
similar to those of \vb.  Spectra of the two first objects display
a range of H$_{\alpha}$ emission profile intensities, strong absorptions
among high level lines, and variable (possibly regular) light curves.
Although the resolution of these spectra is sufficient to show at
least strong forbidden emission lines, none are visible. 
If many of these stars are members of the group exemplified by \va~
and \vb, it will indicate that their properties are not isolated to
a particular evolutionary age and/or angular momentum state. This is a 
possibility that we cannot yet dismiss.

\end{itemize}

\section{Acknowledgments}
REM acknowledges support by Fondecyt grant 1070705, the Chilean 
Center for Astrophysics FONDAP 15010003 and  from the BASAL
Centro de Astrof\'isica y Tecnologias Afines (CATA) PFB--06/2007.
We thank G. Pietrzy\'nski and Dr. Darek Graczyk for his help with the OGLE database and observations.
We thank Daniela Barr\'ia for her help with the reductions of LCO data. We
also express our appreciation to Dr. D. J. Lennon for his suggestion that 
led to our working hypothesis of colliding winds in our program objects.
Our FUSE work was supported by NASA Grant NNX07AC75G to the
Catholic University of America. 
\\

\label{lastpage}
\end{document}